\documentclass[aps,preprint,showpacs,amsmath,amssymb]{revtex4}
\usepackage{graphicx}
\usepackage{longtable}
\usepackage{dcolumn}  
\usepackage{lscape,ulem}
\usepackage{footnote}
\usepackage{color}
\definecolor{navyblue}{rgb}{0.3,0.3,1}
\definecolor{purple}{rgb}{0.6,0,0.5}
\usepackage[colorlinks=true, pdfstartview=FitV, linkcolor=purple, citecolor=
blue,urlcolor=navyblue]{hyperref}
\usepackage[caption=false]{subfig}
\usepackage{subfig}

\input{epsf}
\def\be{\begin{equation}} \def\ee{\end{equation}} \def\bea{\begin{eqnarray}}
\def\eea{\end{eqnarray}}

\def\beq{\begin{equation}}
\def\eeq{\end{equation}}
\def\beqa{\begin{eqnarray}}
\def\eeqa{\end{eqnarray}}

\usepackage{ulem,soul}

\begin{document}

\author{Thamirys de Oliveira}
\affiliation{Departamento de F\'{\i}sica - CFM - Universidade Federal de Santa Catarina, \\ 
Florian\'opolis - SC - CP. 476 - CEP 88.040 - 900 - Brazil \\ email: thamirys.oliveira@posgrad.ufsc.br}

\author{D\'ebora P. Menezes}
\affiliation{Departamento de F\'{\i}sica - CFM - Universidade Federal de Santa Catarina, \\ 
Florian\'opolis - SC - CP. 476 - CEP 88.040 - 900 - Brazil \\ email: debora.p.m@ufsc.br}

\author{Marcus B. Pinto}
\affiliation{Departamento de F\'{\i}sica - CFM - Universidade Federal de Santa Catarina, \\ 
Florian\'{o}polis - SC - CP. 476 - CEP 88.040 - 900 - Brazil \\ email: marcus.benghi@ufsc.br}

\author{Francesca Gulminelli}
\affiliation{CNRS and ENSICAEN, UMR6534, LPC, \\ 
14050 Caen c\'edex, France \\ email:
gulminelli@lpccaen.in2p3.fr }

\title{The role of strangeness and isospin in low density expansions of hadronic matter}

\begin{abstract}
We compare relativistic mean field models with their low
density expansion counterparts used to mimic non-relativistic models by consistently expanding the baryonic scalar density  in powers of the baryonic number density up to ${\cal O}(13/3)$,  which goes two orders beyond the order considered in previous works. We show that, due to the non-trivial density dependence of the Dirac mass, the convergence of the expansion is very slow, and the validity of the non-relativistic approximation {\bf is} questionable even at subsaturation densities. In order to analyze the roles played by strangeness and isospin we consider $n-\Lambda$ and $n-p$ matter separately. Our results indicate that these degrees of freedom play quite different roles in the expansion mechanism and $n-\Lambda$ matter can be better described by low density expansions than $n-p$ matter in general.

\end{abstract}

\maketitle

\section{Introduction}

The properties of nuclear structure as well as the behavior of the nuclear matter equation of state (EOS) have been described over the years with a high degree of precision by phenomenological models. Two main categories of such models exist, namely the non-relativistic Hamiltonian-based functionals, and the relativistic Lagrangian-based ones.

To give a single example of their performance, the most sophisticated present empirical density functionals have attained an accuracy on nuclear mass reproduction  below 0.5 MeV \cite{goriely2013}, that is comparable to the best direct fits of mass tables \cite{WS42014}. Although present relativistic functionals cannot achieve yet this degree of accuracy it is important to mention that  a rms deviation as low as 1.1 MeV was recently obtained within the density dependent meson coupling scheme \cite{artega2016}.

Given this impressive predictive power, one may expect that the extrapolation to nuclear matter properties should give consistent, reliable and model-independent results. 
Indeed, with the continuous development of phenomenological models and the increasing quality of experimental and observational data in the last decade, the uncertainty interval associated to the empirical parameters of the equation of state,
is progressively shrinking \cite{Tsang2012,Dutra2012,Lattimer2013,Dutra2014, Dutra2016}. These parameters give the first coefficients of a Taylor expansion of the energy functional around  the saturation density of symmetric matter, $\rho_0$,   and allow for a complete description of the equation of state
close to  this particular density value \cite{BaoAn_PR}. 

However, the average value of the empirical parameters  shows some systematic differences depending on the fact that the constraints are reproduced using  non-relativistic \cite{Tsang2012,Dutra2012} or relativistic \cite{Dutra2014} functionals; consequently, the predictions for nuclear matter and astrophysical observables also differ \cite{Dutra2008,Fortin2016}. 
This fact is particularly striking if we consider that these differences do not only concern the high density part of the EOS, which requires an important extrapolation from the density region where experimental information exists; systematic differences exist also at densities below \cite{Dutra2008} or around \cite{Fortin2016} saturation, where in principle a relativistic theory should converge towards the non-relativistic limit \cite{Boguta1983}, and moreover a phenomenological functional is by definition strongly constrained by the imposed data fit. 

A particular interesting aspect concerns instability properties, which determine the nuclear liquid-gas phase transition as well as the neutron star crust-core transition. Relativistic and non-relativistic models significantly differ in their qualitative predictions for the spinodals and binodals in asymmetric nuclear matter  \cite{Providencia:2007dp,Dutra2008}. The same is true for the more exotic neutron-$\Lambda$ mixture: a strangeness driven phase transition is observed in a large portion of the parameter space \cite{Fran02} when a Skyrme-based non-relativistic functional is used, while no instability was found within the relativistic mean-field in refs.\cite{Oertel2015,James2017}. This qualitative difference is observed in spite of the fact  that the same constraints are applied to the two classes of models,
and a liquid-gas phase transition at low density exists in both cases \cite{Fran03,James2016}. 

These facts suggest that, independent of the numerical value of the parameters and coupling constants,  the density dependence of the functional is qualitatively different in relativistic and non-relativistic models even at very low densities.

 One conceptual difference comes from the fact that non-relativistic Skyrme based functionals  are complemented by density dependent terms with non-integer powers of density, which effectively simulate many-body effects and cannot be derived from an underlying effective interaction. Such terms  bring correlations among the nuclear empirical parameters and are not present in the relativistic formulation \cite{Khan2012}.
Another source of difference comes from the effective masses, which are systematically lower in relativistic models, due to the strong cancellation between the large scalar and vector potentials. Because of the well-know correlation between the effective mass at saturation and the spin-orbit splitting \cite{Furnstahl1998}, a fit on nuclear masses systematically produces low effective masses unless tensor coupling terms are added in the effective Lagrangian \cite{Rufa1988} or energy dependence is included in the effective masses \cite{Vretenar2002,Antic2015}. 
Finally, the nature itself of the Dirac effective masses implies a very complex implicit density dependence due to the scalar field coupling. This is deeply different from the explicit density dependence of the Landau mass which renormalizes the kinetic term in a non-relativistic formulation. It is therefore possible that even in the low density classical limit, the functional dependence of the relativistic energy density, might be too complex to be obtained from a low density expansion with a small number of parameters.

To progress on these issues, in this paper we develop a systematic low density expansion of the scalar density in powers of the baryonic density, and deduce the corresponding low density expansion of the original relativistic mean-field (RMF) energy functional.  The expansion technique  was originally developed  in  Ref. \cite{Serot1997}, and later applied to the study of asymmetric nuclear matter in Refs. \cite{Margueron:2007jc, Providencia:2007dp}.  

It is interesting to observe that a satisfactory description of RMF models can be obtained if a density expansion is performed around saturation density, as recently proposed in Ref.\cite{functional}. However, the polynomial expansion of Ref.\cite{functional} is an empirical prescription allowing one to describe different families of models within a unique flexible functional, and does not correspond to a non-relativistic limit of the RMF model. On the contrary, the low density expansion can also be viewed as a relativistic expansion in the parameter $k_F/m^*$ \cite{Serot1997}, and therefore allows to quantify the deviation between a relativistic model and its non-relativistic limit. 

Here, our first goal is to consider $n-\Lambda$ baryonic matter in order to analyze how strangeness  affects the low density expansion. To the best of our knowledge such a study has not been carried out before despite its importance regarding, e.g., strangeness driven phase transitions.

Our second  goal is to re-examine the isospin dependence of the nuclear functional by performing a systematic expansion which incorporates higher order contributions, {which should be able to naturally improve many of the results and circumvent some of the problems found in Refs. \cite{Margueron:2007jc, Providencia:2007dp}, which will be discussed  in future sections.}

With respect to the results found in Ref. \cite{ Providencia:2007dp}, we remark that the expansion performed here represents a more  consistent approach since in that paper the nucleon effective masses was taken as being the exact result while in the present application they   are expanded to the relevant perturbative order. 

In the ideal case where the expansion series is fully resummed, the corresponding functional is identical to the RMF  by construction. However, a non-relativistic functional is characterized by a finite number of parameters, corresponding to the truncation of the expansion to a finite order. The deviations observed between the original RMF and the expansion truncated at a finite order thus represent the intrinsic difference between relativistic and non-relativistic formulations, due to the specific functional dependence of the Dirac effective mass. 

The amount of the deviation will obviously depend on the number of terms retained, on the baryonic density interval of the comparison, and on the specific form of the Lagrangian. 
In this work, the convergence of the expansion is studied both for $n-\Lambda$ and $n-p$ matter, and we especially focus on the determination of the number of orders needed to recover the correct behavior of the spinodal borders. 

We show that the convergence of the energy density is relatively fast if we limit ourselves to the sub-saturation regime, in agreement with the results of Ref.\cite{Margueron:2007jc}. However, a very high number of terms is needed to get a convergence of the energy per particle and the chemical potentials, inducing strong and qualitative differences on the instability properties between the complete functional and its non-relativistic approximation.
Surprisingly, the slow convergence of the series is kept even if we consider models with density dependent couplings.

For the analyses of the $n-\Lambda$ matter, a linear RMF model is used
with the inclusion of the usual scalar and vector fields related to
the $\sigma$ and $\omega$ mesons plus the scalar and vector strange
fields, associated with the $\sigma^*$ and $\phi$ mesons, as in 
\cite{schaffner1993,rafael2008,James2016}. As protons are excluded
from this analyses, the vector-isovector  field
corresponding to the $\rho$ meson is not considered.

As for the $n-p$ case, a non-linear RMF model is used and all the
fields that can affect the effective mass and the energy functional
are considered, i.e., the scalar-isoscalar $\sigma$ meson, the
vector-isoscalar $\omega$ meson, the vector-isovector  $\rho$ meson
and the scalar-isovector $\delta$ meson \cite{Dutra2014}.

This paper is organized as follows: in Section II, the complete
formalism used for $n-\Lambda$ matter within the RMF model is reviewed and
the expansion expressions are explicitly written, the
numerical results are displayed and commented. For $n-p$ matter the
formalism is introduced  and the results are presented in
Section III. The final conclusions are summarized in Section IV.

\section{$n-\Lambda$ matter}

To describe meson mediated baryonic interactions between the neutron and the (strange) $\Lambda$ baryon  let us consider the relativistic Lagrangian density

\begin{equation}
{\cal L}_{n \Lambda} = {\cal L}_0^b + {\cal L}_0^m + {\cal L}_i^Y \;\;,
\end{equation}
where the free baryonic term is given by
\begin{equation}
{\cal L}_0^b = \sum_{b=n}^\Lambda {\bar \psi}_b ( i \gamma_\mu \partial^\mu - M_b)\psi_b \;\;.
\end{equation}
The free mesonic term is represented by
\begin{equation}
{\cal L}_0^m = \frac{1}{2}[( \partial_\mu \sigma)^2 - m_\sigma^2 \sigma^2] + \frac{1}{2}[( \partial_\mu \sigma^*)^2 - m_{\sigma^*}^2 (\sigma^*)^2] -\frac{1}{2}\left [ \frac{1}{2} \Omega_{\mu \nu} \Omega^{\mu \nu} - m_\omega^2 \omega_\mu^2 \right ] -\frac{1}{2}\left [ \frac{1}{2} \Phi_{\mu \nu} \Phi^{\mu \nu} - m_\phi^2 \phi_\mu^2 \right ]
\end{equation}
represent $\sigma$ and $\sigma^*$ are scalar mesons while $\omega$ and
 $\Phi$  represent vector mesons. The interactions between baryons are mediated by  meson exchanges through Yukawa vertices as described by 
\begin{equation}
{\cal L}_i^Y =  \sum_{b=n}^\Lambda {\bar \psi}_b \left (g_{\sigma b} \sigma   - g_{\omega b}  \gamma_\mu \omega^\mu  \right )\psi_b +
{\bar \psi}_\Lambda \left (g_{\sigma^* \Lambda} \sigma^* - g_{\phi
    \Lambda}\gamma_\mu \phi^\mu \right )\psi_\Lambda, 
\label{intlnl}
\end{equation}
where $g_{i \Lambda}=\chi_{i \Lambda} g_{in}$, with the mesons denoted
by $i=\sigma, \omega, \sigma^*, \phi$ while $\chi_{i \Lambda}$ is a numerical factor to be defined below. 
The last term in Eq. (\ref {intlnl}) shows that hyperonic and nucleonic degrees of freedom are treated in an asymmetric fashion since the former type of baryon can self interact via two extra channels (mediated by the strange mesons $\sigma^*$ and $\phi$).

Another approach that has been shown to provide a good description
both in nuclear matter and finite nuclei applications is the one in
which the couplings between baryons and mesons depend on the medium density.
The original prescription \cite{tw} was improved while the
applications were developed, and in the present work we choose the
parametrization of the
density dependent hadronic model with $\delta$ mesons, known as
DDH$\delta$ \cite{ddhd,ddhdr}, which was shown to satisfy many experimental
\cite{Dutra2014} and astrophysical constraints \cite{Dutra2016}. As
already said, neither $\rho$ nor $\delta$ mesons are included in the
study of $n-\Lambda$ matter, but they will be used next when $n-p$
matter is investigated. Within this approach, the couplings in
Eq.(\ref{intlnl}) are replaced by
\begin{eqnarray}
g_{\sigma n}\to\Gamma_{\sigma n}(\rho),\quad g_{\omega
  n}\to\Gamma_{\omega n}(\rho),\quad
g_{\sigma^* n}\to\Gamma_{\sigma^* n} (\rho)\quad\mbox{and}\quad 
g_{\phi n}\to\Gamma_{\phi n}(\rho),
\label{ddcouplnp}
\end{eqnarray}
where
\begin{eqnarray}
\Gamma_{i n}(\rho) &=& \Gamma_{i n}(\rho_0)f_i(x),\quad\mbox{with}\quad
f_i(x) = a_i\frac{1+b_i(x+d_i)^2}{1+c_i(x+d_i)^2}
\quad\mbox{and}\quad x=\rho/\rho_0,
\label{gamadefault}
\end{eqnarray}
with $\rho_0$ representing the nuclear saturation density while
$a_i,b_i,c_i$ and $d_i$ are constants that fit some DBHF calculations and finite
nuclei properties, and 
$\Gamma_{i \Lambda}=\chi_{i \Lambda} \Gamma_{in}$.

\subsection{Mean Field Results at zero temperature}
Using standard mean field approximation techniques the relevant results may be expressed in terms of scalar and vector densities as well as mean field values. We can now define the total baryonic scalar density $\rho_s = \langle {\bar \psi} \psi \rangle = {\rho_s}_n + {\rho_s}_\Lambda$ where each  baryon contributes with
\begin{equation}
{\rho_s}_b= \langle {\bar \psi}_b \psi_b \rangle = 2 \int_0^{k_{F_b}} \frac {d^3 {\bf  k}}{(2\pi)^3} \frac {M^*_b}
{[{\bf k}^2 + {M_b^*}^2]^ {1/2}}\;\;.
\label{densityS}
\end{equation}
Analogously one may define the total (number) density as $\rho=\langle {\psi}^+ \psi \rangle=\rho_n + \rho_\Lambda$ where each individual contribution reads
\begin{equation}
\rho_{b}= \langle {\psi_b}^+ \psi_b \rangle = 2 \int_0^{k_{F_b}} \frac
{d^3 {\bf  k}}{(2\pi)^3} =
\frac {k_{F_b}^3}{3\pi^2} \;\;,
\label{density}
\end{equation}
where $k_{F_b}$ is the Fermi momentum
while $M_b^*$ represents the baryon effective mass to be defined below. 
 The Euler-Lagrange equations can now be solved by applying the mean field approximation upon imposing
translational invariance and rotational symmetry of infinite nuclear 
matter. Within this framework  the mesonic equations of motion are then readily obtained and the corresponding
 mean field values read
\begin{equation}
\sigma  = \sum_{b=n}^\Lambda \frac {g_{\sigma b}}{m_\sigma^2} {\rho_s}_b\;\;,
\label{elsigma}
\end{equation}
\begin{equation}
\sigma^*  = \frac {g_{\sigma^* \Lambda}}{m_{\sigma^*}^2} {\rho_s}_\Lambda\;\;,
\label{elsigma*}
\end{equation}
\begin{equation}
\omega^0  =\sum_{b=n}^\Lambda \frac {g_{\omega b}}{m_\omega^2}   \rho_{b}\;\;,
\label{elomega0}
\end{equation}
and 
\begin{equation}
\phi^0  =\frac {g_{\phi \Lambda}}{m_\phi^2}
\rho_{\Lambda} .
\label{elphi0}
\end{equation}

The effective baryonic masses are given by
\begin{equation}
M_n^* = M_n - \frac{g_{\sigma n}}{m_\sigma^2} ( g_{\sigma n} {\rho_s}_n+ g_{\sigma \Lambda} {\rho_s}_\Lambda)\;\;,
\end{equation}
and
\begin{equation}
M_\Lambda^* = M_\Lambda - \frac{g_{\sigma n} g_{\sigma \Lambda }}{m_\sigma^2} {\rho_s}_n - \left( 
\frac{g_{\sigma \Lambda}^2}{m_\sigma^2} +\frac{g_{{\sigma^*} \Lambda}^2}{m_{\sigma^*}^2} \right ) {\rho_s}_\Lambda\;\;.
\end{equation}
For a particular baryon, the  effective chemical potential is given by
\begin{equation}
\mu_b^*=(k_{F_b}^2+{M_b^*}^2)^{1/2} \;,
\end{equation}
so that the individual chemical potentials become
\begin{equation}
\mu_n= \mu_n^*  +   \frac{g_{\omega n} }{m_\omega^2} ( g_{\omega n} \rho_{n}+ g_{\omega \Lambda} \rho_{\Lambda})\;\;,
\label{mun}
\end{equation}
and
\begin{equation}
\mu_\Lambda = \mu_\Lambda^*  + \frac{g_{\omega n} g_{\omega \Lambda }}{m_\omega^2} \rho_{n} + \left( 
\frac{g_{\omega \Lambda}^2}{m_\omega^2} +\frac{g_{{\phi} \Lambda}^2}{{m_\phi}^2} \right ) \rho_{\Lambda}\;\;.
\label{muL}
\end{equation}

At this point it is important to recall that a first order phase transition is signaled by an instability in the
mean-field thermodynamic total energy density, ${\cal E}$, which can be depicted
in terms of a spinodal area whose determination  can be achieved by defining the so called curvature matrix, $\mathbf{C}$. This matrix is associated to 
the  scalar function ${\cal E}$ at a point denoted by 
$P\in\left(\rho_{n}\times \rho_{\Lambda}\right)$ while its elements
are  the second derivatives of ${\cal E}$ with 
respect to each independent variable, $\rho_b$. In our case the curvature matrix
is just a $2 \times 2$ matrix with elements \cite{Fran02, Margueron:2007jc,James2016}:
\begin{equation}
C_{b b^\prime}= \frac{\partial ^{2}{\cal E} \left( \rho_{b},\rho_{b^\prime}\right) }{\partial \rho_{b}\partial \rho_{b^\prime}}
=\left(\frac{\partial \mu_b}{\partial \rho_{b^\prime}}\right),
\label{curvature}
\end{equation}
where $b,b^\prime=n,\Lambda$, whose eigenvalues and eigenvectors acquire a
geometric meaning if $P$ is a critical point. We can solve their roots
explicitly and they read:
\begin{equation}
\lambda _{1 }=\frac{1}{2}\left(\mathtt{Tr}\left( \mathbf{C}\right)+\sqrt{\mathtt{Tr}\left( \mathbf{C}\right) ^{2}-4\mathtt{Det}\left( \mathbf{C}\right) }\right)
\end{equation}
and
\begin{equation}
\lambda _{2 }=\frac{1}{2}\left( \mathtt{Tr}\left( \mathbf{C}\right)-\sqrt{\mathtt{Tr}\left( \mathbf{C}\right) ^{2}-4\mathtt{Det}\left( \mathbf{C}\right) }\right),
\end{equation}
where $\mathtt{Det}\left( \mathbf{C}\right)=\lambda _{1}\lambda _{2}$
and $\mathtt{Tr}\left( \mathbf{C}\right)=\lambda _{1}+\lambda _{2}$.
To build the spinodal area, the modulus of the negative eigenvalue is used for each possible combination of neutron and $\Lambda$ densities.  For more details, we refer the reader to the papers just cited as well as to Ref. \cite{DG}. As  Eq. (\ref {curvature}) shows, the knowledge of the chemical potentials given by Eqs. (\ref {mun}) and (\ref{muL}) is necessary for the determination of the spinodal region. 

For our purposes it is convenient to 
 express the energy density in terms of the scalar and baryon number densities as
$$
{\cal E}({\rho_s}_b,\rho_b) = \sum_{b=n}^\Lambda \left ( \frac {3}{4}
  \rho_b \epsilon_b + \frac{M_b^*}{4} {\rho_s}_b \right ) + \frac{1}{2
  m_\sigma^2} \left ( \sum_{b=n}^\Lambda  g_{\sigma b} {\rho_s}_b
\right )^2 +\frac{g_{{\sigma^*} \Lambda}^2}{2 m_{\sigma^*}^2}
{\rho_s}_\Lambda^2 \nonumber 
$$
\begin{equation}
+ \frac{1}{2 m_\omega^2} \left ( \sum_{b=n}^\Lambda  g_{\omega b} \rho_{b} \right )^2+
 \frac{g_{{\phi} \Lambda}^2}{2{m_\phi}^2} \rho_{\Lambda}^2 \;,
\label{energy}
\end{equation}
where 
\begin{equation}
\epsilon_b= [(3 \pi^2 \rho_b)^{2/3}+{M_b^*}^2]^{1/2} \;,
\label{singlepart}
\end{equation}
is the single particle energy density.

In all equations above, when the DDH$\delta$ model is used, the
substitution given in Eq.(\ref{ddcouplnp}) has to be done. The
chemical potentials, however, are modified by a rearrangement term
given by \cite{Oertel2015}:

\begin{equation}
\Sigma^R_{n \Lambda} (\rho)=\sum_b \left(
\frac{\partial\Gamma_{\omega  b}}{\partial\rho_b} \omega^0\rho_b
+\frac{\partial\Gamma_{\phi^0  b}}{\partial\rho_b} \phi^0\rho_b
-\frac{\partial\Gamma_{\sigma b}}{\partial\rho_b} \sigma \rho_{s b}
-\frac{\partial\Gamma_{\sigma^* b}}{\partial\rho_b}\sigma^*\rho_{s b} 
\right),
\end{equation}
which, after the equations of motion given in Eqs.(\ref{elsigma})-(\ref{elphi0}) are used to replace the fields by the corresponding densities, can be rewritten as:
\begin{eqnarray}
\Sigma^R_{n \Lambda} (\rho)&=&\sum_{b=n}^{\Lambda} \sum_{b^\prime=n}^{\Lambda} \left [\left(
\frac{\partial\Gamma_{\omega  b}}{\partial\rho} \right )\frac{\Gamma_{\omega  b^\prime}}{m_\omega^2} \rho_b \rho_{b^\prime}
-\left (\frac{\partial\Gamma_{\sigma b}}{\partial\rho}\right ) \frac{\Gamma_{\sigma  b^\prime}}{m_\sigma^2} {\rho_s}_b {\rho_s}_{b^\prime}
\right] \nonumber \\
&+&\left( \frac{\partial\Gamma_{\phi \Lambda  }}{\partial\rho}\right )\frac{\Gamma_{\phi \Lambda}}{m_\phi^2} \rho_\Lambda^2
-\left ( \frac{\partial\Gamma_{\sigma^* \Lambda }}{\partial\rho}\right ) \frac{\Gamma_{\sigma^* \Lambda}}{m_{\sigma^*}^2} {\rho_s}_\Lambda^2 \;.
\end{eqnarray}
Then, the chemical potentials become:
\begin{equation}
\mu_n = \mu_n^* +  \frac{\Gamma_{\omega n} }{m_\omega^2} ( \Gamma_{\omega n}
\rho_{n}+ \Gamma_{\omega \Lambda} \rho_{\Lambda}) + \Sigma_{n \Lambda}^R(\rho)\;\;,
\end{equation}
and
\begin{equation}
\mu_\Lambda = \mu_\Lambda^* + \frac{\Gamma_{\omega n} \Gamma_{\omega \Lambda }}{m_\omega^2} \rho_{n} + \left( \frac{\Gamma_{\omega \Lambda}^2}{m_\omega^2} +\frac{\Gamma_{{\phi}
    \Lambda}^2}{{m_\phi}^2} \right ) \rho_{\Lambda} + \Sigma_{n \Lambda}^R(\rho)\;\;.
\end{equation}

\subsection{Low density expansion}

To make contact with non relativistic model, which depends only on
$\rho$, one may perform the integral in Eq. (\ref {densityS}) and then expand the result in powers of $\rho$ \cite{Serot1997}.  In principle, using a computing software such as {\it Mathematica}$^{\textregistered}$ one may easily perform such an expansion to arbitrarily high orders. Here, as already emphasized we expand the scalar density up to order-$\rho^{13/3}$ obtaining
\begin{equation}
{\rho_s}_b= \rho_b + \frac{c_1}{{M_b^*}^2} \rho_b^{5/3} + \frac{c_2} {{M_b^*}^4}
\rho_b^{7/3} + \frac{c_3}{{M_b^*}^6} \rho_b^{3} + 
\frac{c_4}{{M_b^*}^8} \rho_b^{11/3}
+\frac{c_5}{{M_b^*}^{10}} \rho_b^{13/3} + {\cal O}(\rho_b^{15/3}) \;\;.
\label {exprhos}
\end{equation}
where
$c_1= - 3(3\pi^2)^{2/3}/10$, $c_2= 9(3\pi^2)^{4/3}/56$, $c_3=-15(3\pi^2)^{2}/144$, $c_4=105(3\pi^2)^{8/3}/1408$, and $c_5=-189(3\pi^2)^{10/3}/3328$. 


The next step is to substitute this expansion into Eq. (\ref {energy})
and then consistently re-expand to the desired order in  $\rho_b$. 
Also it is convenient to define  the dimensionful  (${\rm eV}^{-2}$)
quantities 
$f_n = g_{\sigma n}^2/m_\sigma^2$, $f_{n \Lambda}= g_{\sigma n} g_{\sigma \Lambda} /m_\sigma^2$, $f_\Lambda= ( g_{\sigma \Lambda}^2/m_\sigma^2 + g_{\sigma^* \Lambda}^2/
m_\sigma^2)$, $h_n = g_{\omega n}^2/m_\omega^2$, $h_{n \Lambda}= g_{\omega n} g_{\omega \Lambda} /m_\omega^2$, $h_\Lambda= g_{\omega \Lambda}^2/m_\omega^2$ and $r_\Lambda = g_{\phi \Lambda}^2/m_\phi^2$. 

Recalling that we consider ${\rho_s}_b$ as well as ${\cal E}$ expanded up to order-$\rho^{13/3}$ and   inspecting  Eq. (\ref {energy}) one concludes that   the effective masses must be expanded at least up to order-$\rho^{10/3}$ as implied, e.g., by the term proportional to ${\rho_s}_b M^*_b$. Then, to this particular order the effective masses can be written as

\begin{eqnarray}
M_\Lambda^* &=& M_\Lambda - (f_\Lambda \rho_\Lambda + f_{n \Lambda} \rho_n) - c_1 \left ( \frac{ f_\Lambda \rho_\Lambda^{5/3}}{M_\Lambda^2} + \frac {f_{n \Lambda} \rho_n^{5/3}}{M_n^2} \right ) 
- c_2  \left ( \frac{ f_\Lambda \rho_\Lambda^{7/3}}{M_\Lambda^4} + \frac {f_{n \Lambda} \rho_n^{7/3}}{M_n^4} \right )
\nonumber  \\
&-&2 c_1 \left( \frac {f_\Lambda^2 \rho_\Lambda^{8/3}}{M_\Lambda^3} + \frac{f_{n \Lambda} f_\Lambda \rho_\Lambda^{5/3} \rho_n}{M_\Lambda^3}+ \frac{f_{n \Lambda}^2 \rho_n^{5/3} \rho_\Lambda}{M_n^3}+ \frac{ f_{n \Lambda} f_n \rho_n^{8/3}}{M_n^3} \right ) -
c_3 \left ( \frac{f_\Lambda \rho_\Lambda^3}{M_\Lambda^6} + \frac{f_{ n \Lambda} \rho_n^3}{M_n^6} \right ) \nonumber \\ 
&-&2c_1^2\left(\frac{f_{n \Lambda}^2 \rho_n^{5/3}\rho_{\Lambda}^{5/3}}{M_n^3 M_{\Lambda}^2}+ \frac{f_{ n \Lambda}f_n\rho_n^{10/3}}{M_n^5} + \frac{f_{\Lambda}^2 \rho_{\Lambda}^{10/3}}{M_{\Lambda}^5} + \frac{f_{\Lambda} f_{n\Lambda} \rho_n^{5/3}\rho_{\Lambda}^{5/3}}{M_n^2 M_{\Lambda}^3}\right) \nonumber \\
&-&4c_2\left(\frac{f_{\Lambda}^2\rho_{\Lambda}^{10/3}}{M_{\Lambda}^5} + \frac{f_{\Lambda} f_{n\Lambda}\rho_{\Lambda}^{7/3}\rho_n}{M_{\Lambda}^5} + \frac{f_{n\Lambda}^2\rho_n^{7/3}\rho_{\Lambda}}{M_n^5} + \frac{f_{n\Lambda}f_n\rho_n^{10/3}}{M_n^5} \right)  +  {\cal O}(\rho_b^{11/3}),
\label{effml}
\end{eqnarray}
and
\begin{eqnarray}
M_n^* &=& M_n - ( f_n \rho_n + f_{n \Lambda} \rho_\Lambda) - c_1 \left ( \frac{ f_n \rho_n^{5/3}}{M_n^2} + \frac {f_{n \Lambda} \rho_\Lambda^{5/3}}{M_\Lambda^2} \right ) 
- c_2  \left ( \frac{ f_n \rho_n^{7/3}}{M_n^4} + \frac {f_{n \Lambda} \rho_\Lambda^{7/3}}{M_\Lambda^4} \right ) \nonumber \\
&-&2 c_1 \left( \frac {f_n^2 \rho_n^{8/3}}{M_n^3} + \frac{f_{n \Lambda} f_n \rho_n^{5/3} \rho_\Lambda}{M_n^3}+ \frac{f_{n \Lambda}^2 \rho_\Lambda^{5/3} \rho_n}{M_\Lambda^3}+ \frac{ f_{n \Lambda} f_\Lambda \rho_\Lambda^{8/3}}{M_\Lambda^3} \right )-
c_3 \left ( \frac{f_n \rho_n^3}{M_n^6} + \frac{f_{ n \Lambda} \rho_\Lambda^3}{M_\Lambda^6} \right ) \nonumber \\ 
&-&2c_1^2\left(\frac{f_{n \Lambda}^2 \rho_n^{5/3}\rho_{\Lambda}^{5/3}}{M_n^2 M_{\Lambda}^3}+ \frac{f_{ n \Lambda}f_{\Lambda}\rho_{\Lambda}^{10/3}}{M_{\Lambda}^5} + \frac{f_n^2 \rho_n^{10/3}}{M_n^5} + \frac{f_n f_{n\Lambda} \rho_n^{5/3}\rho_{\Lambda}^{5/3}}{M_n^3 M_{\Lambda}^2}\right) \nonumber \\
&-&4c_2\left(\frac{f_n^2\rho_n^{10/3}}{M_n^5} + \frac{f_n f_{n\Lambda}\rho_n^{7/3}\rho_{\Lambda}}{M_n^5} + \frac{f_{n\Lambda}^2\rho_{\Lambda}^{7/3}\rho_n}{M_{\Lambda}^5} + \frac{f_{n\Lambda}f_{\Lambda}\rho_{\Lambda}^{10/3}}{M_{\Lambda}^5} \right)  +  {\cal O}( \rho_b^{11/3}).
\label{effmn}
\end{eqnarray}

In terms of the bare baryonic masses, the order-$\rho^{13/3}$ scalar densities become
\begin{eqnarray}
{\rho_s}_n&=& \rho_n + c_1 \frac{\rho_n^{5/3}}{M_n^2} + c_2 \frac{\rho_n^{7/3}}{M_n^4} 
+2c_ 1 \frac{[f_{n \Lambda} \rho_\Lambda \rho_n^{5/3} + f_n \rho_n^{8/3}]}{M_n^3} + c_3 \frac{\rho_n^3}{M_n^6} \nonumber \\
&+& 2c_1^2 \frac{f_{n\Lambda} \rho_\Lambda^ {5/3} \rho_n^ {5/3}}{M_\Lambda^ 2 M_n^ 3} +
4 c_2 \frac{ f_{n\Lambda} \rho_\Lambda \rho_n^ {7/3}}{M_n^ 5} + 2(c_1^ 2+2c_2) \frac{f_n \rho_n^ {10/3}}{M_n^ 5}\nonumber \\
&+& 3c_1 \frac{f_{n\Lambda}^ 2 \rho_\Lambda^ 2 \rho_n^ {5/3}}{M_n^ 4} + 6 c_1 \frac {f_n f_{n\Lambda} \rho_\Lambda \rho_n^ {8/3}}{M_n^ 4} + 3 c_1\frac{ f_n^2 \rho_n^ {11/3}}{M_n^ 4} + c_4 \frac {\rho_n^{11/3}}{M_n^ 8}\nonumber \\
&+& 2c_1 c_2 \frac{f_{n\Lambda} \rho_\Lambda^{7/3}\rho_n^{5/3}}{M_\Lambda^4 M_n^3} +
4c_1 c_2 \frac{ f_{n\Lambda}\rho_\Lambda^{5/3} \rho_n^ {7/3}}{M_\Lambda^ 2 M_n^ 5} +
6c_3 \frac {f_{n\Lambda}\rho_\Lambda \rho_n^ 3}{M_n^7} + 6(c_1 c_ 2+c_3) \frac{f_n \rho_n^4}{M_n^ 7} \nonumber \\
&+& \dfrac{10c_1^2}{M_n^6}(f_n^2 \rho_n^{13/3} + f_n f_{n\Lambda}\rho_\Lambda \rho_n^{10/3}) + \dfrac{6c_1^2}{M_n^4 M_\Lambda^2}(f_n f_{n\Lambda} \rho_n^{8/3} \rho_\Lambda^{5/3} + f_{n\Lambda}^2 \rho_n^{5/3} \rho_\Lambda^{8/3})\nonumber \\
&+& \dfrac{4c_1^2}{M_n^3 M_\Lambda^3}(f_\Lambda f_{n\Lambda} \rho_n^{5/3} \rho_\Lambda^{8/3} + f_{n\Lambda}^2 \rho_n^{8/3} \rho_\Lambda^{5/3}) + 10c_2 \dfrac{\rho_n^{7/3}}{M_n^6}(f_n \rho_n + f_{n\Lambda} \rho_\Lambda)^2 + c_5 \dfrac{\rho_n^{13/3}}{M_n^{10}} \nonumber \\
&+& {\cal O}(\rho^{14/3}) \;,
\end{eqnarray}
and 
\begin{eqnarray}
{\rho_s}_\Lambda &=&\rho_\Lambda + c_1 \frac{\rho_\Lambda^{5/3}}{M_\Lambda^2} + c_2 \frac{\rho_\Lambda^{7/3}}{M_\Lambda^4} + 2c_ 1 \frac{[f_{n \Lambda} \rho_n \rho_\Lambda^{5/3} + f_\Lambda \rho_\Lambda^{8/3}]}{M_\Lambda^3} + c_3 \frac{\rho_\Lambda^3}{M_\Lambda^6} \nonumber \\
&+& 2c_1^2 \frac{f_{n\Lambda} \rho_\Lambda^ {5/3} \rho_n^ {5/3}}{M_n^ 2 M_\Lambda^ 3} +
4 c_2 \frac{ f_{n\Lambda} \rho_n \rho_\Lambda^ {7/3}}{M_\Lambda^ 5} + 2(c_1^ 2+2c_2) \frac{f_\Lambda \rho_\Lambda^ {10/3}}{M_\Lambda^5}\nonumber \\
&+& 3c_1 \frac{f_{n\Lambda}^ 2 \rho_n^2 \rho_\Lambda^ {5/3}}{M_\Lambda^4} + 6 c_1 \frac {f_\Lambda f_{n\Lambda} \rho_n \rho_\Lambda^ {8/3}}{M_\Lambda^4} + 3 c_1\frac{ f_\Lambda^2 \rho_\Lambda^ {11/3}}{M_\Lambda^ 4} + c_4 \frac {\rho_\Lambda^{11/3}}{M_\Lambda^ 8}\nonumber \\
&+& 2c_1 c_2 \frac{f_{n\Lambda} \rho_n^{7/3}\rho_\Lambda^{5/3}}{M_n^4 M_\Lambda^3} +
4c_1 c_2 \frac{ f_{n\Lambda}\rho_n^{5/3} \rho_\Lambda^ {7/3}}{M_n^ 2 M_\Lambda^ 5} +
6c_3 \frac {f_{n\Lambda}\rho_n \rho_\Lambda^ 3}{M_\Lambda^7} + 6(c_1 c_ 2+c_3) \frac{f_\Lambda \rho_\Lambda^4}{M_\Lambda^ 7} \nonumber \\
&+& \dfrac{10c_1^2}{M_\Lambda^6}(f_\Lambda^2 \rho_\Lambda^{13/3} + f_\Lambda f_{n\Lambda}\rho_n \rho_\Lambda^{10/3}) + \dfrac{6c_1^2}{M_\Lambda^4 M_n^2}(f_\Lambda f_{n\Lambda} \rho_\Lambda^{8/3} \rho_n^{5/3} + f_{n\Lambda}^2 \rho_\Lambda^{5/3} \rho_n^{8/3})\nonumber \\
&+&\dfrac{4c_1^2}{M_n^3 M_\Lambda^3}(f_n f_{n\Lambda} \rho_\Lambda^{5/3} \rho_n^{8/3} + f_{n\Lambda}^2 \rho_\Lambda^{8/3} \rho_n^{5/3}) + 10c_2 \dfrac{\rho_\Lambda^{7/3}}{M_\Lambda^6}(f_\Lambda \rho_\Lambda + f_{n\Lambda} \rho_n)^2 + c_5 \dfrac{\rho_\Lambda^{13/3}}{M_\Lambda^{10}}\nonumber \\
&+& {\cal O}(\rho^{14/3}) \;.
\end{eqnarray}

One can then define the power expansion for the energy density as
\begin{equation}
{\cal E}(\rho_b)= \sum_{b=n}^\Lambda [{\cal K}_b+M_b \rho_b] + {\cal E}_{pot}(\rho_b)\;\;, \label{eq:non-rel-like}
\end{equation}
where ${\cal K}_b= -c_1 \rho_b^{5/3}/M_b$ is the kinetic energy   while the dynamics is described by the  power series ${\cal E}_{pot}(\rho_b) = \sum_{k=6}^{13} {\cal E}_{k/3}$
whose individual  terms read:

\begin{equation}
{\cal E}_{2}=\frac{1}{2} \left [ r_\Lambda \rho_\Lambda^2 +\left ( h_\Lambda \rho_\Lambda^2 + 2 h_{\Lambda n} \rho_n \rho_\Lambda + h_n \rho_n^2 \right )   - \left ( f_\Lambda \rho_\Lambda^2 + 2 f_{\Lambda n} \rho_n \rho_\Lambda + f_n \rho_n^2 \right ) \right ] \;\;,
\end{equation}

\begin{equation}
{\cal E}_{7/3}= - \frac{c_2}{3} \left ( \frac{\rho_\Lambda^{7/3}}{M_\Lambda^3} + 
\frac{\rho_n^{7/3}}{M_n^3} \right )\;\;,
\end{equation}

\begin{equation}
{\cal E}_{8/3}= -c_1 \left [ f_\Lambda \frac{\rho_\Lambda^{8/3}}{M_\Lambda^2} +f_{\Lambda n}\left ( \frac{\rho_\Lambda^{5/3} \rho_n}{M_\Lambda^2} + \frac{\rho_n^{5/3} \rho_\Lambda}{M_n^2}\right ) + f_n \frac{\rho^{8/3}_n}{M_n^2} \right ] \;\;,
\end{equation}

\begin{equation}
{\cal E}_{3}= - \frac {c_3}{5} \left ( \frac{\rho_\Lambda^3}{M_\Lambda^5} + \frac{\rho_n^3}{M_n^5}\right ) \;\;,
\end{equation}

\begin{equation}
{\cal E}_{10/3}=-c_1^2 \left \{ \frac{16}{7} \left ( f_\Lambda \frac{\rho_\Lambda^{10/3}}{M_\Lambda^4} +
f_n \frac{\rho_\Lambda^{10/3}}{M_n^4} \right ) + f_{ \Lambda n}\left [ \frac{ \rho_n^{5/3} \rho_\Lambda^{5/3}}{M_n^2 M_\Lambda^2} + \frac {25}{14} \left ( \frac{\rho_n \rho_\Lambda^{7/3}}{M_\Lambda^4}+\frac{\rho_\Lambda \rho_n^{7/3}}{M_n^4}
\right )\right ] \right \}\;,
\end{equation}

\begin{eqnarray}
{\cal E}_{11/3}&=& -\frac{1}{7} c_4 \left ( \frac {\rho_\Lambda^{11/3}}{M_\Lambda^7} + \frac {\rho_n^{11/3}}{M_n^7} \right ) - c_ 1 \left [ f_\Lambda^2\frac {\rho_\Lambda^{11/3}}{M_\Lambda^3} +f_n^2 \frac {\rho_n^{11/3}}{M_n^3}+f_{n\Lambda}^2 \left ( \frac {\rho_\Lambda^{5/3} \rho_n^2}{M_\Lambda^3} +  \frac {\rho_n^{5/3} \rho_\Lambda^2}{M_n^3}  \right )\right]\nonumber \\
&-&2c_1 \left[  f_{n\Lambda}\left ( f_\Lambda \frac
    {\rho_\Lambda^{8/3} \rho_n}{M_\Lambda^3}+ f_n \frac {\rho_n^{8/3}
    \rho_\Lambda}{M_n^3}  \right ) \right] \;,
\end{eqnarray}

\begin{eqnarray}
{\cal E}_{4}&=& -4 c_1 c_2 \left( \frac{f_n \rho_n^4}{M_n^6} + \frac{f_{n \Lambda} \rho_n^{7/3}\rho_{\Lambda}^{5/3}}{M_n^4 M_{\Lambda}^2} + \frac{f_{n\Lambda}\rho_{\Lambda}^{7/3}\rho_n^{5/3}}{M_{\Lambda}^4 M_n^2} + \frac{f_\Lambda \rho_{\Lambda}^4}{M_{\Lambda}^6} \right) \nonumber \\
&-& c_3 \left( \frac{f_n \rho_n^4}{M_n^6} +
    \frac{f_{n\Lambda} \rho_n^3 \rho_\Lambda}{M_n^6} +
    \frac{f_{n\Lambda} \rho_{\Lambda}^3 \rho_n}{M_\Lambda^6} +
    \frac{f_\Lambda \rho_{\Lambda}^4}{M_{\Lambda}^6} \right) \;,
\end{eqnarray}
and
{\small
\begin{eqnarray}
{\cal E}_{13/3}&=&-2c_1^2 \left[
                   f_nf_{n\Lambda}\left(\frac{\rho_n^{8/3}\rho_\Lambda^{5/3}}{M_n^3
                   M_{\Lambda}^2}+\frac{\rho_n^{10/3}\rho_\Lambda}{M_n^5}\right)+
                   f_{\Lambda}f_{n\Lambda}\left(\frac{\rho_n^{5/3}\rho_\Lambda^{8/3}}{M_n^2
                   M_{\Lambda}^3}+\frac{\rho_n
                   \rho_{\Lambda}^{10/3}}{M_{\Lambda}^5}\right) \right . \nonumber \\
&+& \left . f_{n\Lambda}^2\left(\frac{\rho_n^{8/3}\rho_\Lambda^{5/3}}{M_n^2 M_{\Lambda}^3}+\frac{\rho_n^{5/3}\rho_\Lambda^{8/3}}{M_n^3 M_{\Lambda}^2}\right)\right] 
- (c_1^2 +c_2) \left[
    f_n^2\left(\frac{\rho_n^{13/3}}{M_n^5}\right)+f_{\Lambda}^2\left(\frac{\rho_{\Lambda}^{13/3}}{M_{\Lambda}^5}\right)\right] \nonumber\\
&-& 2c_2 \left[ 2f_nf_{n\Lambda}\left(\frac{\rho_n^{10/3}\rho_\Lambda}{M_n^5}\right)+f_{n\Lambda}^2\left(\frac{\rho_n^{7/3}\rho_\Lambda^{2}}{M_n^5}+\frac{\rho_n^{2}\rho_\Lambda^{7/3}}{M_\Lambda^5}\right) +2f_{\Lambda}f_{n\Lambda}\left(\frac{\rho_n \rho_\Lambda^{10/3}}{M_\Lambda^5}\right)\right]\nonumber \\
& -&\frac{c_5}{9}
\left(\frac{\rho_n^{13/3}}{M_n^9}+\frac{\rho_\Lambda^{13/3}}{M_\Lambda^9}\right) .
\end{eqnarray}}
It is important to stress that much simpler functional forms are supposed in the case of phenomenological non-relativistic functionals, both in the case of zero range Skyrme and finite range effective interactions. As a consequence, a mapping between the two formalisms can only be done in the density region where Eq.(\ref{eq:non-rel-like}) can be approximated by retaining a limited number of terms.

The explicit expansion of the rearrangement term, necessary for the analyses of the DDH$\delta$ model, up to order-$\rho^{13/3}$, is given in the appendix.

\subsection{Numerical Results}

We start by studying the convergence of the expansion in $n-\Lambda$
matter. For this purpose, we use the linear parametrization of the
Walecka model also used in Ref. \cite{James2016} and fix the couplings
such that the constraints imposed in Ref. \cite{James2017} are obeyed.
In this parametrization the meson masses are given by 
$m_\sigma=550$ MeV, $m_\omega=783$ MeV,
$m_{\sigma^*}=980$ MeV and $m_\phi=1020$ MeV and the couplings by
$g_{\sigma n}=\sqrt{91.64}$ and $g_{\omega n}=\sqrt{136.2}$.
Among the possible choices given in Table I of Ref. \cite{James2017}, we
have chosen $\chi_{\sigma \Lambda}=0.5$, $\chi_{\omega \Lambda}=0.522$,
$\chi_{\sigma^*  \Lambda}=0.5$ and $\chi_{\phi \Lambda}=0.612$.
 Whenever the density dependent model is used, the constants are
  fixed following Refs. \cite{Liu2002,MP2004} although, as already said, the
  $\rho$ and $\delta$ mesons are not necessary in the study of
  $n-\Lambda$ matter. Finally, let us also define the $\Lambda$ fraction as 
$Y_\Lambda=\rho_\Lambda/\rho$.

\begin{figure}[h]
\begin{tabular}{cc}
\subfloat[]
{\includegraphics[width=0.5\textwidth]{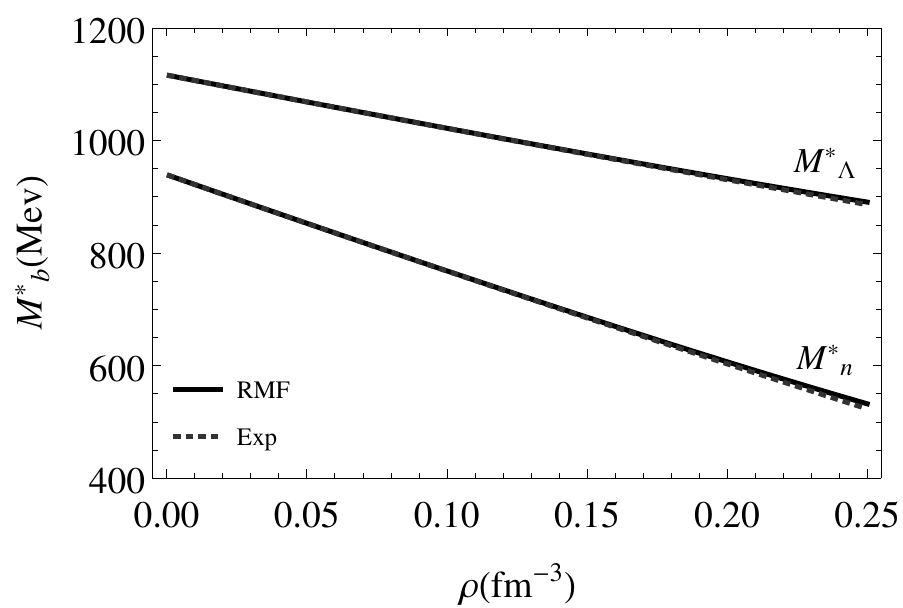}} &
\subfloat[]
{\includegraphics[width=0.5\textwidth]{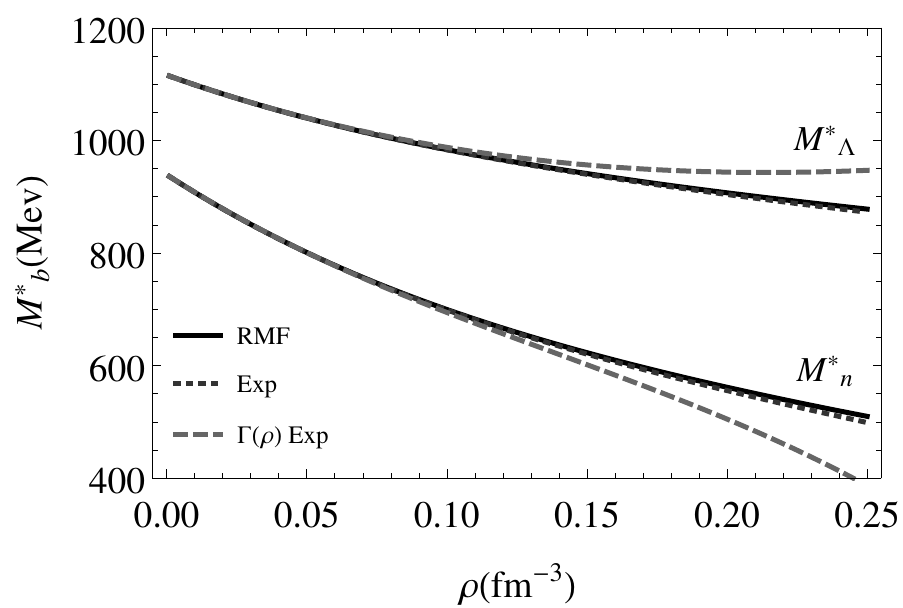}} \\
\end{tabular}
\caption{(Color online) Effective masses, up to $O(\rho^{10/3})$, with $\Lambda$ fraction $Y_\Lambda=0.5$ compared with  RMF (a)  and  DDH $\delta$ model (b). For comparison purposes, the right panel  also shows the result obtained by performing the $\rho$ expansion in the DD couplings, $\Gamma(\rho)$}
\label{fig_mass_nl}
\end{figure}

We first plot the $\Lambda$ and neutron effective masses
obtained from the RMF model and from the expansion given respectively in 
Eq.(\ref{effml}) and (\ref{effmn}) for both models with constant
couplings and with density dependent ones in Fig.\ref{fig_mass_nl}. 

\begin{figure}[h]
	\begin{tabular}{cc}
		\subfloat[]
		{\includegraphics[width=0.5\textwidth]{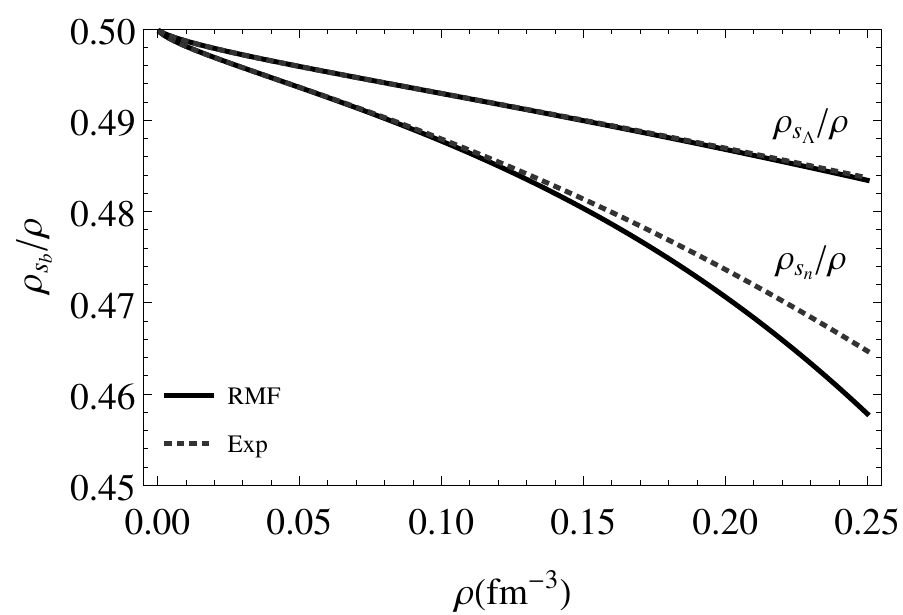}} &
		\subfloat[]
		{\includegraphics[width=0.5\textwidth]{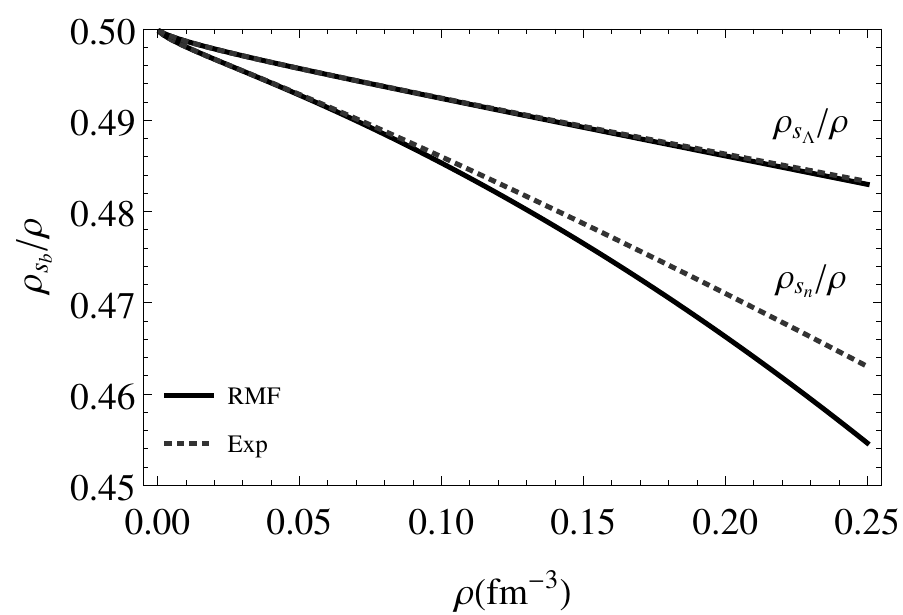}} \\
	\end{tabular}
\caption{(Color online) Normalized scalar densities,  up to $O(\rho^{13/3})$, with $Y_\Lambda=0.5$ compared with  RMF (a)  and  DDH $\delta$ model (b).}
\label{fig_rhos}
\end{figure}  

As one can see,   up to  $\rho \approx 0.25 {\rm fm}^{-3}$, the agreement is very good when terms of order-$\rho^{10/3}$ are considered. It is worth mentioning that we have also investigated the inclusion of terms up to order-$\rho^{12/3}$ (not shown) noting that the results shown in Fig. \ref {fig_mass_nl} remain very stable.
For completeness, we have also addressed the possibility of consistently expanding the DD couplings, $\Gamma(\rho)$, 
and the results are shown with dotted lines in Fig. 1(b) which indicates that the (full) proposed functional forms given
  in Eqs. (\ref{ddcouplnp}) already take the density dependence correctly. Therefore, we shall consider only the complete DD couplings in all the subsequent evaluations. 
  
As already emphasized,  the scalar density plays an essential role regarding our low density evaluations, so it is important to look at its power expansion in some detail. In Fig. \ref{fig_rhos} we plot the order-$\rho^{13/3} $ expansion predictions for this quantity  showing that the  approximation performs remarkably well for the ${\rho_s}_\Lambda$ case in the full density range considered in that figure but starts to deviate  around $0.1\, {\rm fm}^{-3}$ for the
neutrons. This suggests that the extra interaction, mediated by the $\sigma^*$ meson, plays a major role as far as the convergence of the scalar density power series is concerned. We point out that the same degree of agreement is observed in the density behavior of the energy density (not shown).

\begin{figure}[h]
\begin{tabular}{cc}
\subfloat[]
{\includegraphics[width=0.5\textwidth]{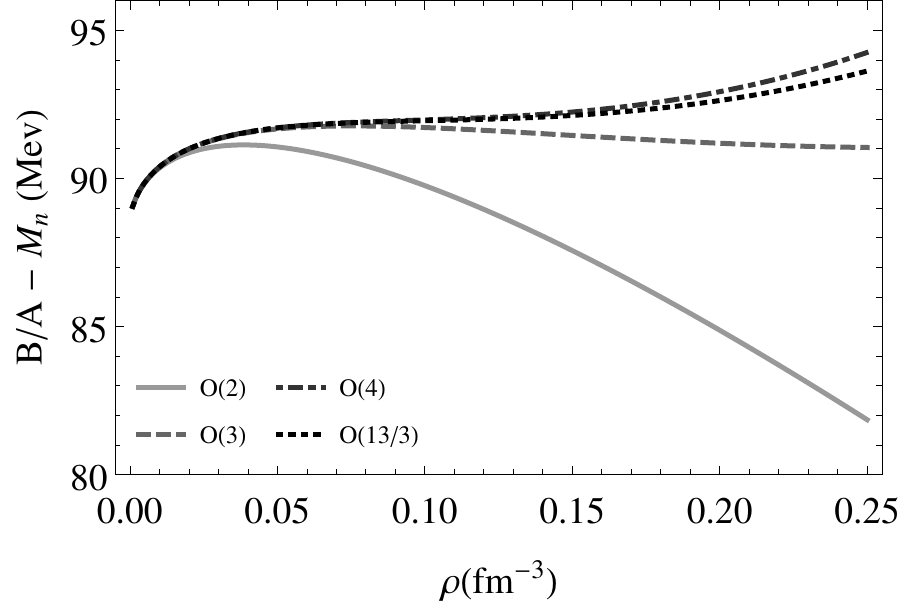}} &
\subfloat[]
{\includegraphics[width=0.5\textwidth]{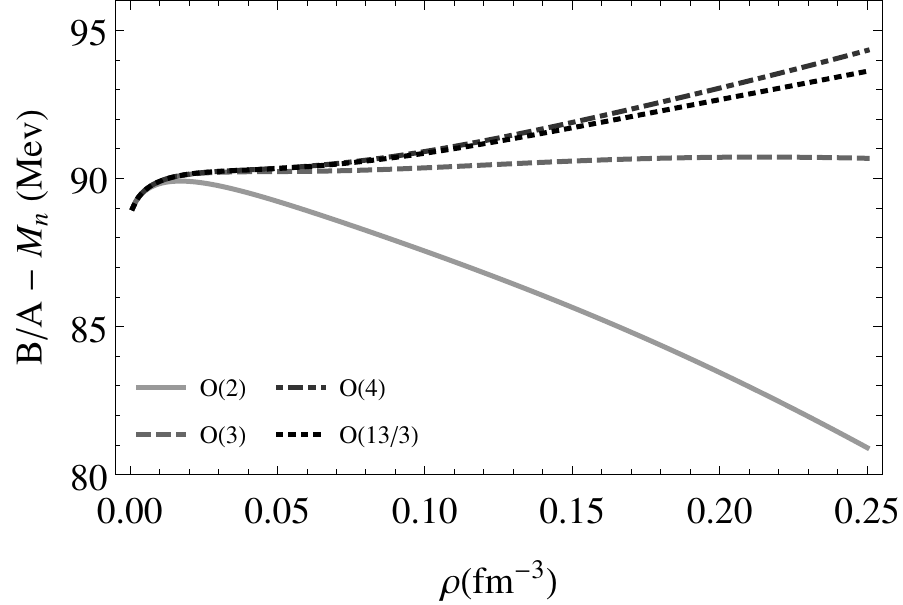}} \\
\end{tabular}
\caption{(Color online)Order by order contributions, up to order-$\rho^{13/3}$, to the binding energy expansion when $Y_\Lambda=0.5$ for the  RMF model (a) and for the  DDH$\delta$ model (b).}
\label{fig_order_nl}
\end{figure}

In order to examine the convergence properties of the expansion, and appreciate the importance of the relativistic effects on the EOS,
we now look at the (order by order) binding energy expansion defined as
$B/A -M_n {\rm (MeV)} = {\cal E}/\rho -M_n {\rm (MeV)}$ when $Y_\Lambda= 0.5$. Each curve
in Fig.\ref{fig_order_nl}
contains the mentioned order plus the previous ones so that the ${\cal  O}(\rho^{13/3})$ curve
 includes the whole expansion up to this
order and this is the result we use next, to compare with the RMF calculation.

\begin{figure}[h]
\begin{tabular}{cc}
\subfloat[]
{\includegraphics[width=0.5\textwidth]{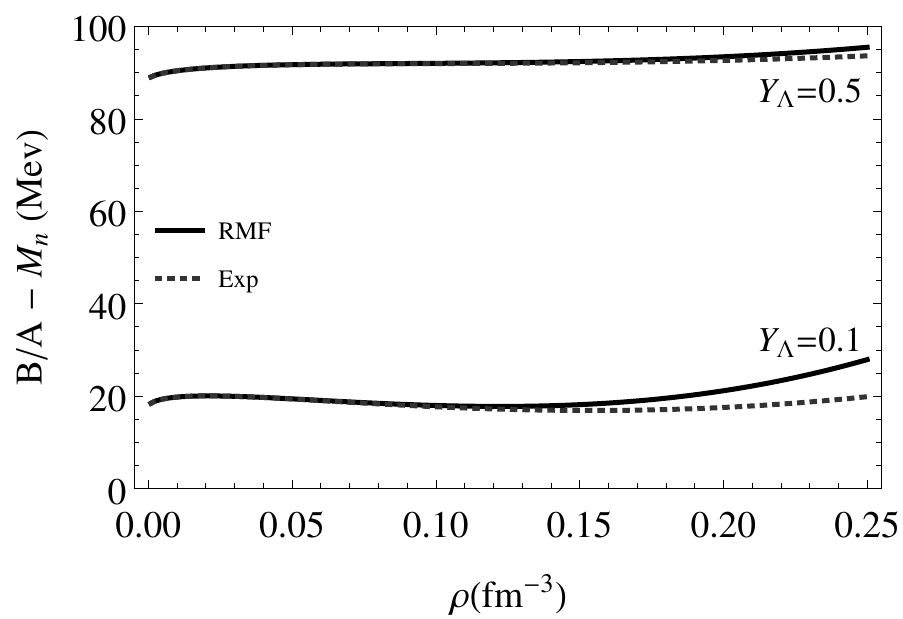}} &
\subfloat[]
{\includegraphics[width=0.5\textwidth]{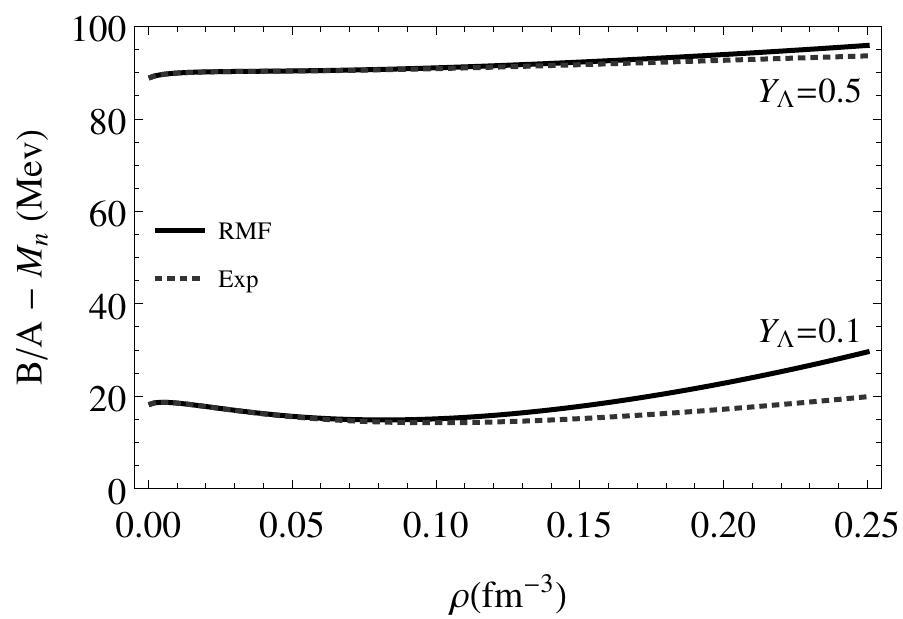}} \\
\end{tabular}
\caption{(Color online)Binding energy for $\Lambda$ fractions $Y_\Lambda=0.1$ (bottom curves) and $Y_\Lambda=0.5$
  (top curves) with the RMF model(left panel) and the  DDH$\delta$ model (right panel). The expansion considers contributions up to order-$\rho^{13/3}$}
\label{fig_binding_nl}
\end{figure}

In Fig. \ref{fig_binding_nl}, we plot the binding energy obtained with
the RMF models and their low density expansions for two particular $\Lambda$
fractions: $Y_\Lambda=0.1$ and $Y_\Lambda=0.5$. We can see that huge cancellations take place between neighboring orders and that sizeable deviations are seen in this quantity already around saturation, even if we include all terms up to  ${\cal O}(\rho^{13/3})$. The situation is not drastically changed if density dependent couplings are used as the right panel shows.

\begin{figure}[h]
\begin{tabular}{cc}
\subfloat[]
{\includegraphics[width=0.45\textwidth]{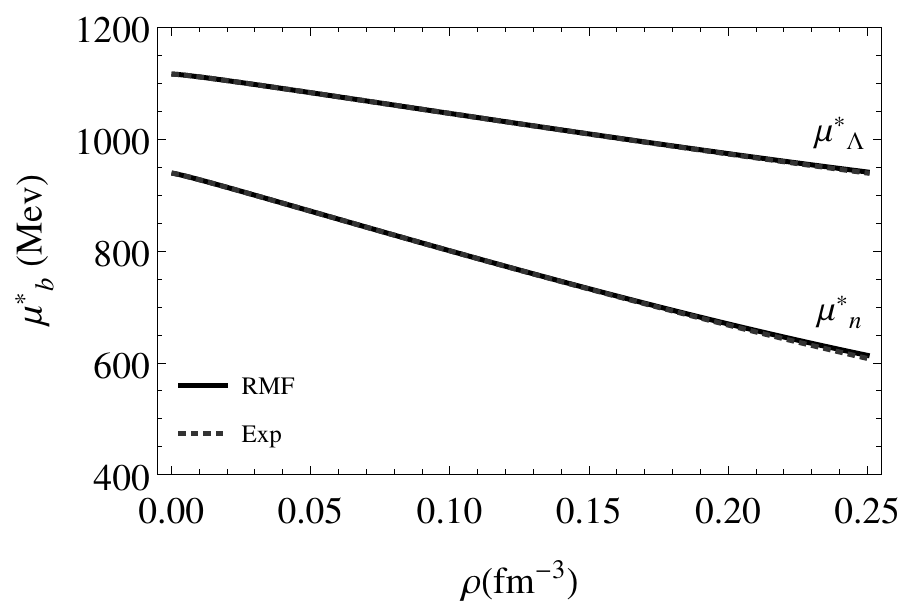}}  &
\subfloat[]
{\includegraphics[width=0.45\textwidth]{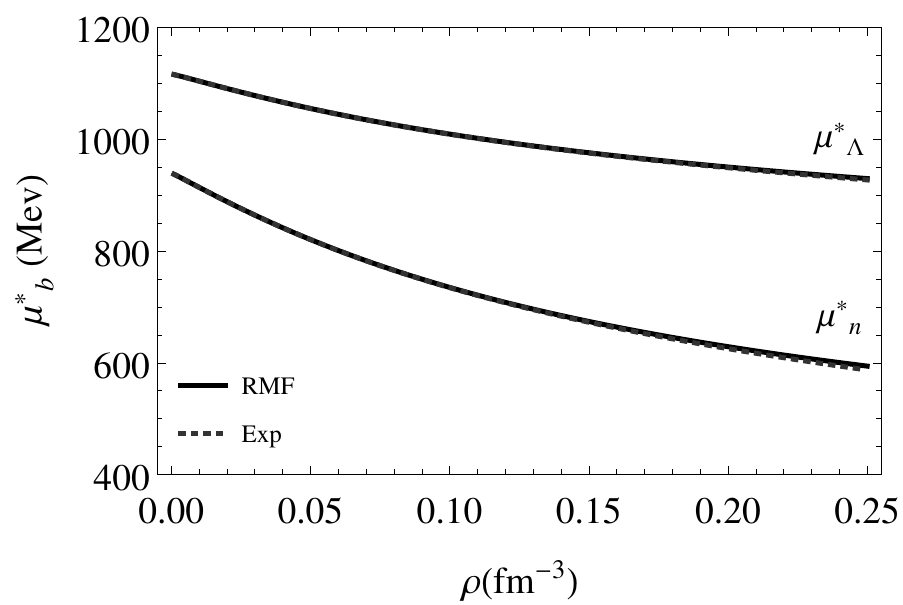}}  \\
\subfloat[]
{\includegraphics[width=0.45\textwidth]{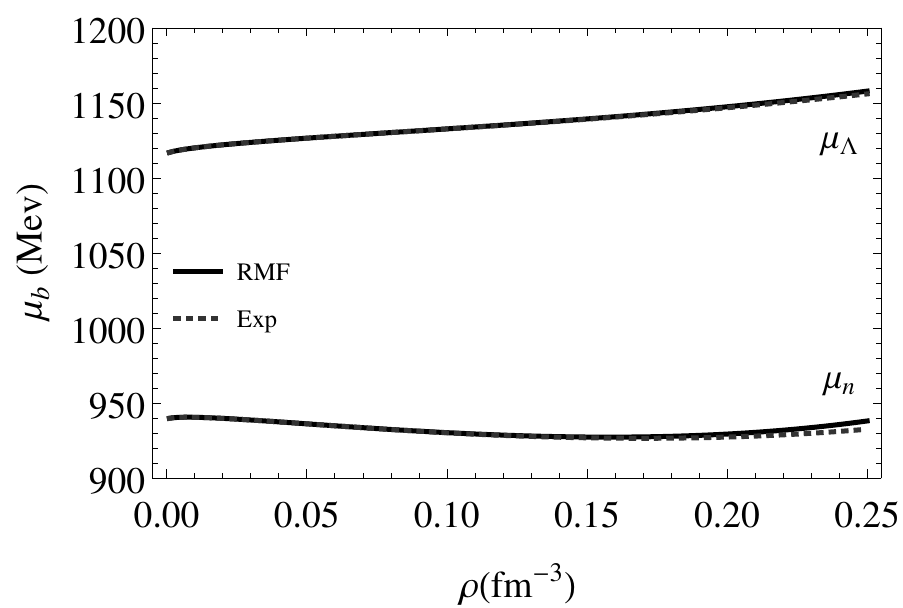}} &
\subfloat[]
{\includegraphics[width=0.45\textwidth]{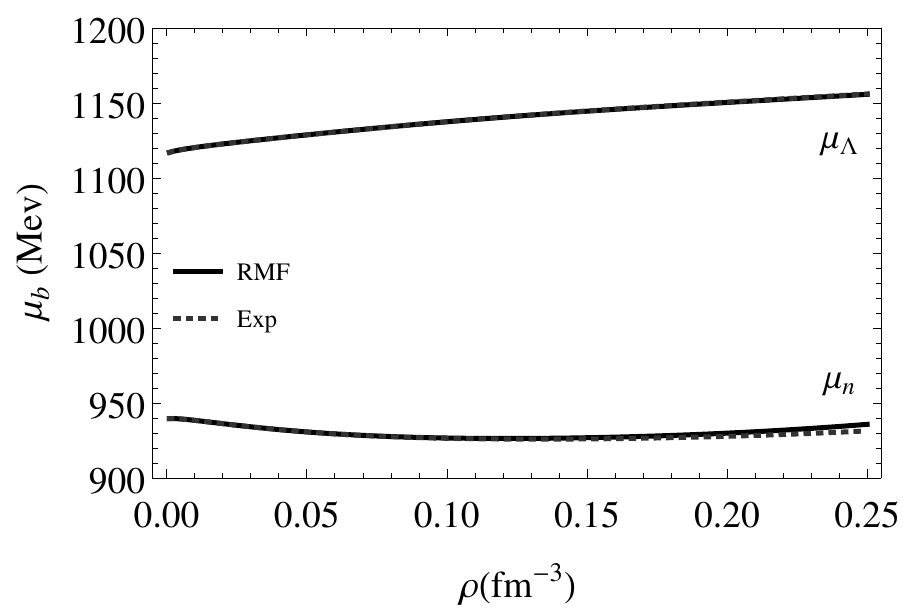}} \\
\end{tabular}
\caption {(Color online) Effective chemical potentials  for the RMF model (a) and the DDH$\delta$
  model (b). chemical potentials for for the RMF model (c) and the DDH$\delta$
  model (d). In both cases the $\Lambda$ fraction was set to $Y_\Lambda=0.5$ while the density expansion was performed up to order-$\rho^{13/3}$.}
\label{fig_nu_mu_nl}
\end{figure}

As stated before, the chemical potentials are important quantities in
the study of instabilities and we next look carefully at the
convergence of their ${\cal O}(\rho^{13/3}$) expansions by comparing, in Fig.\ref{fig_nu_mu_nl},  both the effective and
the chemical potentials with the exact results  considering the
$Y_\Lambda=0.5$ case.

The degree of reproduction of the exact RMF is satisfactory at least
up to saturation density. 

\begin{figure}[h]
\begin{tabular}{cc}
\subfloat[]
{\includegraphics[width=0.5\textwidth]{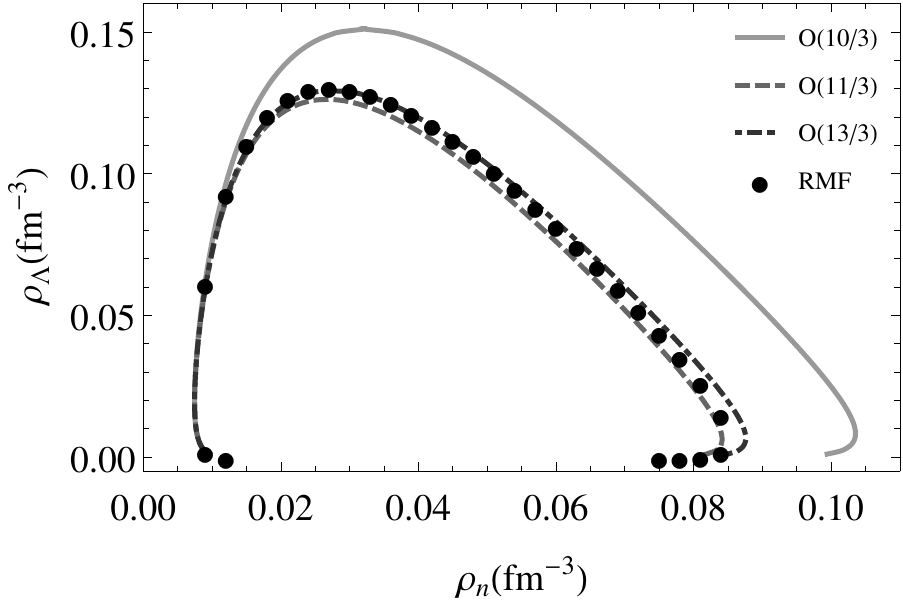}} &
\subfloat[]
{\includegraphics[width=0.5\textwidth]{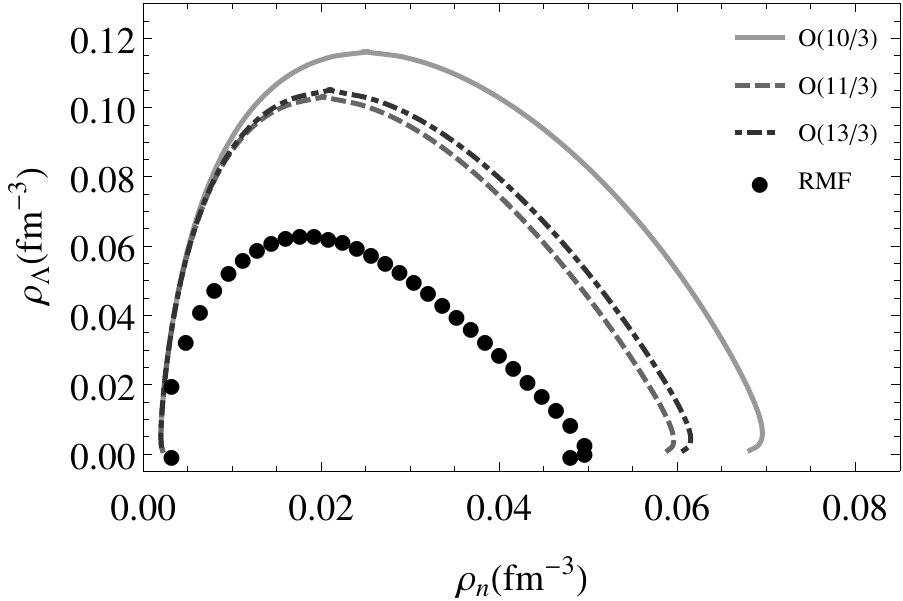}} \\
\end{tabular}
\caption{(Color online) Spinodal sections for (a) RMF and (b) DDH$\delta$  model. The expansion results shown are for orders $\rho^{10/3}$, $\rho^{11/3}$, and $\rho^{13/3}$.}
\label{fig_spinodalnl}
\end{figure}

Finally, if one is interested in studying instabilities at
sub-saturation densities, the spinodal is the quantity of interest. On the left panel of 
 Fig. \ref{fig_spinodalnl}, the spinodal sections, obtained from RMF
 and from the low density expansion at orders  $\rho^{10/3}$, $\rho^{11/3}$, and $\rho^{13/3}$ are plotted.  As one can see the expansion displays a nice oscillatory convergence behavior and the 
 curves practically coincide  at orders $\rho^{11/3}$ and $\rho^{13/3}$. The right panel of the same figure shows that the agreement is less spectacular for the orders considered here. We believe that the  observed 
 deviation is due to the fact that the scalar densities play a more
 important role in density dependent models through the rearrangement term, which is part of the chemical potentials. 
From the above discussion, related to Fig. \ref {fig_rhos}, we recall  that the neutron scalar density
stops converging still at quite low densities. It is then fair to say that the expansion works slightly better for the $n-\Lambda$ model with fixed couplings  than for the density dependent model, with consequences in the spinodal boundary.

\section{$n-p$ matter}

In order to examine isospin effects we now turn our attention to $n-p$ matter whose low density expansion was originally analyzed in Refs. \cite{Margueron:2007jc, Providencia:2007dp} where terms of order-$\rho^{11/3}$ have been considered while here we extend the same expansion to order-$\rho^{13/3}$ so that eventual discrepancies  found in those works may now  be identified as being due to the poor convergence of the lowest order contributions considered. Then, in order to get a better insight on the low density expansion properties it will be instructive to highlight the differences between our (higher order) power series and the ones considered in those seminal works.
One general nonlinear finite range RMF model, introduced by Boguta and Bodmer \cite{boguta} 
 that describes nuclear matter is represented by the following Lagrangian density:

\begin{equation}
{\cal L}_{n p} = {\cal L}_0^b + {\cal L}_0^m + {\cal L}_i^Y + {\cal L}_i^m \;\;.
\label{dl}
\end{equation}
To keep the notation consistent with the one used in the previous $n-\Lambda$ case one can write the free baryonic and mesonic terms as
\begin{equation}
{\cal L}_0^b=\sum_{b=n}^p \overline{\psi}_b(i\gamma^\mu\partial_\mu - M_b)\psi_b ,
\end{equation}
and
\begin{equation}
{\cal L}_0^m = \frac{1}{2}[( \partial_\mu \sigma)^2 - m_\sigma^2 \sigma^2] +
\frac{1}{2}[( \partial_\mu {\vec \delta})^2 - m_\delta^2 {\vec \delta}^2]-\frac{1}{2}\left [ \frac{1}{2} \Omega_{\mu \nu} \Omega^{\mu \nu} - m_\omega^2 \omega_\mu\omega^\mu\right ] -\frac{1}{2}\left [ \frac{1}{2} {\vec B}_{\mu \nu}  {\vec B}^{\mu \nu} - m_\rho^2 \vec{\rho}_\mu\vec{\rho}^\mu \right ]\,,
\end{equation}
respectively.
The Yukawa type of interactions are given by 
\begin{equation}
 {\cal L}_i^Y= g_\sigma\sigma\overline{\psi}_b\psi_b 
- g_\omega\overline{\psi}_b\gamma^\mu\omega_\mu\psi_b 
- \frac{g_\rho}{2} \overline{\psi}_b\gamma^\mu\vec{\rho} _\mu\cdot \vec{\tau}\psi_b
+ g_\delta \overline{\psi}_b \vec{\delta}\cdot \vec{\tau}\psi_b \;,
\end{equation}
while (non linear) self mesonic interactions are described by 
\begin{equation}
 {\cal L}_i^m=- \frac{a}{3}\sigma^3 - \frac{b}{4}\sigma^4 \;.
 \label{Lnl}
 \end{equation}
 Note that contrary to the $n-\Lambda$, analyzed previously,   within this Lagrangian density the baryons do  not necessarily need to be distinguished, when their bare mass difference is neglected, since
they interact through the same 
mesons $j=\sigma,\delta,\omega$, and $\rho$.  
Nevertheless, to be consistent with the previous section as well as for generality reasons we have chosen a notation  which distinguishes $M_p$ and $M_n$ despite the fact that in our numerical evaluations we use the same numerical input for both (bare) masses. Notice also that we have opted to use the notation presented in Ref. \cite{Providencia:2007dp} instead of the one in Ref. \cite{Liu2002} for the couplings and we always take the isospin projection $\tau_3=\pm 1$ respectively for protons and neutrons.
Then,  after applying the mean field procedure for the calculation of the
equations of motion for the meson fields one obtains the following expectation values:
\begin{align}
\sigma &= \frac{1}{m^2_\sigma }(g_\sigma\rho_s - a\sigma^2 - b\sigma^3) ,
\label{sigmaacm}\\
\omega_0 &= \frac{g_\omega}{m_\omega^2}\rho ,
\label{omegaacm}\\
\bar{\rho}_{0(3)} &= \frac{g_\rho}{2 m_\rho^2}\rho_3 ,
\label{rhoacm} \\
\delta_{(3)} &= \frac{g_\delta}{m_\delta^2}\rho_{s3} .
\label{deltaacm}
\end{align}
Once again, to keep in line with the previous notations, we have defined  scalar and vector densities, appropriate to treat the $n-p$ case, as

\begin{align}
\rho_s &=\langle {\bar \psi} \psi \rangle={\rho_s}_p+{\rho_s}_n,\quad
\rho_{s3}=\langle {\bar \psi} \tau_3 \psi \rangle={\rho_s}_p-{\rho_s}_n,
\label{rhos}\\
\rho &= \langle {\psi}^+ \psi \rangle=\rho_p + \rho_n,\quad
\rho_3= \langle {\psi}^+ \tau_3 \psi \rangle=\rho_p - \rho_n.
\label{rho}
\end{align}
where ${\rho_s}_b$ and ${\rho}_b$ can be trivially obtained from Eqs (\ref {densityS}) and (\ref {density}) upon considering  $b=n,p$.  
We can also define the effective nucleon masses as 
\begin{eqnarray}
M_p^*=M_p-g_\sigma\sigma -g_\delta\delta_{(3)}
\qquad\mbox{and}\qquad
M_n^*=M_n -g_\sigma\sigma+g_\delta\delta_{(3)}
\label{effectivemasses}
\end{eqnarray}
and the chemical potentials as
\begin{eqnarray}
\mu_p &=& \mu_p^* +  g_\omega\omega_0 +\frac{g_\rho}{2}\bar{\rho}_{0(3)} ,
\\
\mu_n&=& \mu_n^* + g_\omega\omega_0 -\frac{g_\rho}{2}\bar{\rho}_{0(3)} .
\end{eqnarray}
Note the effect of the meson $\delta$, which splits the effective masses $M_p^*$ and 
$M_n^*$. For symmetric nuclear matter $\delta_{(3)}$ vanishes, since 
$\rho_{s_p}=\rho_{s_n}$, and consequently, $M_p^*=M_n^*$.

For the purpose of the present work, the energy density is required
and reads:
\begin{equation}
{\cal E}({\rho_s}_b,\rho_b) = \sum_{b=n}^p \left ( \frac {3}{4}
  \rho_b \epsilon_b + \frac{M_b^*}{4} {\rho_s}_b \right ) + \frac{1}{2}m^2_\sigma\sigma^2 
+ \frac{a}{3}\sigma^3 + \frac{b}{4}\sigma^4 + \frac{g_\omega^2}{2m^2_\omega}\rho^2 
 +\frac{g_\rho^2}{8 m_\rho^2}\rho_3^2
+ \frac{g_\delta^2}{2 m^2_\delta}\rho_{s3}^2
\label{denerg}
\end{equation}
where the single particle energy, $\epsilon_b$, can be readily obtained from Eq. (\ref {singlepart}) upon considering $b=n,p$.

Once again, we exploit the DDH$\delta$ model for $n-p$ matter. In this case,
the DDH$\delta$ parameterization has the same coupling parameters as 
in Eq.~(\ref{gamadefault}) for the mesons $\sigma$ and $\omega$, but
functions 
\begin{eqnarray}
\Gamma_j(\rho) &=& \Gamma_j(\rho_0)f_j(x),\quad\mbox{with}\quad
f_j(x)=a_je^{-b_j(x-1)}-c_j(x-d_j),
\quad\mbox{and}\quad x=\rho/\rho_0,
\end{eqnarray}
for the isovector mesons $j=\rho,\delta$. In this case, the
rearrangement term reads
\begin{eqnarray}
\Sigma^R_{np}(\rho)=\frac{\partial\Gamma_\omega} {\partial\rho}\omega_0\rho
+\frac{1}{2}\frac{\partial\Gamma_\rho }{\partial\rho}\bar{\rho}_{0(3)}\rho_3
-\frac{\partial\Gamma_\sigma}{\partial\rho}\sigma\rho_s
-\frac{\partial\Gamma_\delta}{\partial\rho}\delta_{(3)}\rho_{s3},
\end{eqnarray}
which can be rewritten in terms of the densities after the equations of motion are used to replace the mesons fields
as 
\begin{eqnarray}
\Sigma^R_{np}(\rho)=\left(\frac{\partial\Gamma_\omega} {\partial\rho}\right ) \frac{\Gamma_\omega}{m_\omega^2} \rho^2
+\frac{1}{4}\left (\frac{\partial\Gamma_\rho }{\partial\rho}\right )\frac{\Gamma_\rho}{m_\rho^2}  \rho_3^2
-\left(\frac{\partial\Gamma_\sigma}{\partial\rho}\right ) \sigma\rho_s
-\left (\frac{\partial\Gamma_\delta}{\partial\rho}\right) \frac{\Gamma_\delta}{m_\delta^2} \rho_{s3}^2\,,
\end{eqnarray}
whose order-$\rho^{13/3}$ is explicitly given in the appendix.

Within  the $n-p$ case the chemical potentials, for the DDH$\delta$ model read
\begin{eqnarray}
\mu_p&=& \mu_p^*+ \Gamma_\omega\omega_0 +\frac{\Gamma_\rho}{2}\bar{\rho}_{0(3)}
+\Sigma^R_{np},
\\
\mu_n&=& \mu_n^*+ \Gamma_\omega\omega_0 -\frac{\Gamma_\rho}{2}\bar{\rho}_{0(3)}
+ \Sigma^R_{np}.
\end{eqnarray}

\subsection{Low density expansion}

 In previous works \cite{Providencia:2007dp, Margueron:2007jc},
  low density expansions were also calculated for $n-p$ matter, but we
  would like to point out some important differences. 
As already mentioned above, in Ref. \cite{Providencia:2007dp}, the effective mass was taken as the
exact RMF result, not providing a completely consistent picture since the expansion was not carried out in a consistent fashion.  Also,  only the energy density was analyzed in that work and no analysis of the spinodal instabilities was performed. Moreover,
 the expansions performed in Ref. \cite{Margueron:2007jc}  have two orders less than the ones attained here. Nevertheless, the expansion coefficients are in agreement (up to the maximum order considered in Ref. \cite{Margueron:2007jc}). Let us start by defining 
$f_\rho=g_\rho^2/m_\rho^2$, $f_\delta=g_\delta^2/m_\delta^2$, $f_n=g_{\sigma}^2/m_{\sigma}^2$, 
$f_\omega=g_{\omega}^2/m_{\omega}^2$,
$f_{nl}=a/m_{\sigma}^6$, $f_m=2a^2/(g_{\sigma}^2 m_{\sigma}^{10})$,
$f_q=b /m_{\sigma}^8$.

The neutron effective mass can be written as 
\begin{eqnarray}
M_{n}^*&=&M_n - f_{n}\rho + f_{\delta}\rho_3 + f_{nl}\rho^2  - c_1 (f_{n}+f_{\delta}) \dfrac{\rho_n^{5/3}}{M_n^2} - c_1(f_{n}-f_{\delta}) \dfrac{\rho_p^{5/3}}{M_p^2}\nonumber\\ 
&-& c_2 (f_n + f_{\delta})\dfrac{\rho_n^{7/3}}{M_n^4}- c_2 (f_n - f_{\delta})\dfrac{\rho_p^{7/3}}{M_p^4}+ 2 c_1 f_{nl} \rho \left(\dfrac{\rho_n^{5/3}}{M_n^2}+\dfrac{\rho_p^{5/3}}{M_p^2}\right)\nonumber\\ 
&-&2c_1 f_n \left(\dfrac{\rho_n^{5/3}(f_n\rho - f_{\delta}\rho_3)}{M_n^3} + \dfrac{\rho_p^{5/3}(f_n\rho + f_{\delta}\rho_3)}{M_p^3} \right)\nonumber\\ 
&+& 2c_1 f_{\delta}\left(-\dfrac{\rho_n^{5/3}(f_n\rho - f_{\delta}\rho_3)}{M_n^3} + \dfrac{\rho_p^{5/3}(f_n\rho + f_{\delta}\rho_3)}{M_p^3} \right) + (f_q - f_m)\rho^3\nonumber\\ 
&-&c_3 f_n \left(\dfrac{\rho_n^{3}}{M_n^6}+\dfrac{\rho_p^{3}}{M_p^6}\right) +c_3
    f_{\delta}\left(-\dfrac{\rho_n^{3}}{M_n^6}+\dfrac{\rho_p^{3}}{M_p^6}\right) \nonumber \\
&+& (c_1^2+2c_2)\rho_n^{10/3}\left(\dfrac{-2(f_n + f_{\delta})^2}{M_n^5} + \dfrac{f_{nl}}{M_n^4}\right)\nonumber\\ 
&+& 2c_1^2\rho_n^{5/3}\rho_p^{5/3}\left(\dfrac{-(f_n - f_{\delta})^2}{M_n^2 M_p^3}+\dfrac{-f_n^2 + f_{\delta}^2}{M_n^3 M_p^2} +\dfrac{f_{nl}}{M_n^2 M_p^2}\right)\nonumber\\ 
&+&2 c_2\rho_n^{7/3}\rho_p\left(\dfrac{2(-f_n^2 + f_{\delta}^2)}{M_n^5} + \dfrac{f_{nl}}{M_n^4}\right)+2 c_2 \rho_p^{7/3}\rho_n\left(\dfrac{-2(f_n - f_{\delta})^2}{M_p^5} + \dfrac{f_{nl}}{M_p^4}\right)\nonumber\\ 
&+&(c_1^2+2c_2)\rho_p^{10/3}\left(\dfrac{2(-f_n^2 + f_{\delta}^2)}{M_p^5} + \dfrac{f_{nl}}{M_p^4}\right)+{\cal O}(\rho_b^{11/3}) \;,
\end{eqnarray}
while the proton effective mass reads
\begin{eqnarray}
M_{p}^*&=&M_p - f_{n}\rho - f_{\delta}\rho_3 + f_{nl}\rho^2  - c_1 (f_{n}-f_{\delta}) \dfrac{\rho_n^{5/3}}{M_n^2} - c_1(f_{n}+f_{\delta}) \dfrac{\rho_p^{5/3}}{M_p^2}\nonumber\\ 
&-& c_2 (f_n - f_{\delta})\dfrac{\rho_n^{7/3}}{M_n^4}- c_2 (f_n + f_{\delta})\dfrac{\rho_p^{7/3}}{M_p^4}+ 2 c_1 f_{nl} \rho \left(\dfrac{\rho_n^{5/3}}{M_n^2}+\dfrac{\rho_p^{5/3}}{M_p^2}\right)\nonumber\\ 
&-&2c_1 f_n \left(\dfrac{\rho_n^{5/3}(f_n\rho - f_{\delta}\rho_3)}{M_n^3} + \dfrac{\rho_p^{5/3}(f_n\rho + f_{\delta}\rho_3)}{M_p^3} \right)\nonumber\\ 
&+& 2c_1 f_{\delta}\left(\dfrac{\rho_n^{5/3}(f_n\rho - f_{\delta}\rho_3)}{M_n^3} - \dfrac{\rho_p^{5/3}(f_n\rho + f_{\delta}\rho_3)}{M_p^3} \right) + (f_q - f_m)\rho^3\nonumber\\ 
&-&c_3 f_n \left(\dfrac{\rho_n^{3}}{M_n^6}+\dfrac{\rho_p^{3}}{M_p^6}\right) + c3 f_{\delta}\left(\dfrac{\rho_n^{3}}{M_n^6}-\dfrac{\rho_p^{3}}{M_p^6}\right) \nonumber \\
&+&(c_1^2+2c_2)\rho_p^{10/3}\left(\dfrac{-2(f_n + f_{\delta})^2}{M_p^5} + \dfrac{f_{nl}}{M_p^4}\right)\nonumber\\ 
&+&2c_1^2\rho_p^{5/3}\rho_n^{5/3}\left(\dfrac{-(f_n - f_{\delta})^2}{M_p^2 M_n^3}+\dfrac{-f_n^2 + f_{\delta}^2}{M_p^3 M_n^2} +\dfrac{f_{nl}}{M_p^2 M_n^2}\right)\nonumber\\ 
&+&2 c_2 \rho_p^{7/3}\rho_n\left(\dfrac{2(-f_n^2 + f_{\delta}^2)}{M_p^5} + \dfrac{f_{nl}}{M_p^4}\right)+2 c_2 \rho_n^{7/3}\rho_p\left(\dfrac{-2(f_n - f_{\delta})^2}{M_n^5} + \dfrac{f_{nl}}{M_n^4}\right)\nonumber\\ 
&+&(c_1^2+2c_2)\rho_n^{10/3}\left(\dfrac{2(-f_n^2 + f_{\delta}^2)}{M_n^5} + \dfrac{f_{nl}}{M_n^4}\right)+{\cal O}(\rho_b^{11/3})
\end{eqnarray}
From the above expansions for the effective masses  the reader can appreciate a further difference between our  order-$\rho^{10/3}$ expansion and the order-$\rho^{8/3}$ performed in Ref. \cite {Margueron:2007jc}. 
Note that such a  difference arises precisely in the order-$\rho^{3}$ term which is proportional to $f_q=b g_{\sigma}^4/m_{\sigma}^8$ so that, in practice, the quartic $\sigma$ self interaction present in the non linear term, Eq. (\ref{Lnl}), has been completely neglected in Ref.  \cite {Margueron:2007jc}. Moreover, note that at this same order we have $f_m$ which is of order-$a^2$ while in Ref. \cite {Margueron:2007jc} the maximum contribution from the non linear $a \sigma^3$ term is of order-$a$ only.

The $\sigma$ expectation value is given by the non linear self consistent Eq. (\ref {sigmaacm})  so that in Ref.  \cite {Margueron:2007jc} the authors choose to use a low density approximation to solve it by considering the expansion $\sigma = [g_\sigma\rho_s - g_\sigma^2 a \rho_s^2/m_\sigma^2 - {\cal O}(\rho_s^3)]/({m^2_\sigma })$ and thus neglecting the term proportional to $b \sigma^3$. This is an inappropriate course of action if one wishes to describe  higher order contributions so that here  we shall refrain from carrying out further approximations in order to solve Eq. (\ref{sigmaacm}) which is be treated in a fully consistent way according to the order we are working with. 

Let us now expand the nucleonic scalar densities up to order-$\rho^{13/3}$ starting with the one which describes the neutron:
\begin{eqnarray}
{\rho_s}_ n&=& \rho_n +c_1 \dfrac{\rho_n^{5/3}}{M_n^2} + c_2 \dfrac{\rho_n^{7/3}}{M_n^4} +2c_1 \rho_n^{5/3}\left(\dfrac{f_n \rho -f_{\delta}\rho_3}{M_n^3}\right) + c_3 \dfrac{\rho_n^3}{M_n^6}\nonumber\\ 
&+& 2c_1^2 \rho_n^{5/3}f_n\left( \dfrac{\rho_n^{5/3}}{M_n^5}+\dfrac{\rho_p^{5/3}}{M_p^2 M_n^3}\right)+2c_1^2 \rho_n^{5/3}f_{\delta}\left( \dfrac{\rho_n^{5/3}}{M_n^5}-\dfrac{\rho_p^{5/3}}{M_p^2 M_n^3}\right)\nonumber\\ 
&+&4c_2 \rho_n^{7/3}\left(f_n\dfrac{\rho}{M_n^5} - f_{\delta}\dfrac{\rho_3}{M_n^5}\right)+c_4\dfrac{\rho_n^{11/3}}{M_n^8}\nonumber\\ 
&+&c_1\rho_n^{5/3}\left(\dfrac{-2f_{nl}\rho^2}{M_n^3}+\dfrac{3(-f_n \rho+f_\delta(\rho_p-\rho_n))^2}{M_n^4}\right)+ 2c_1 c_2 (f_n - f_{\delta})\frac{\rho_p^{7/3}\rho_n^{5/3}}{M_p^4 M_n^3}\nonumber\\ 
&+& 4c_1 c_2 (f_n - f_{\delta}) \frac{\rho_p^{5/3} \rho_n^ {7/3}}{M_p^ 2 M_n^ 5} +
6c_3 (f_n - f_{\delta}) \frac {\rho_p \rho_n^ 3}{M_n^7} + 6(c_1 c_ 2+c_3) (f_n + f_{\delta}) \frac{\rho_n^4}{M_n^ 7} \nonumber \\
&+& \dfrac{10c_1^2}{M_n^6}[(f_n + f_{\delta})^2 \rho_n^{13/3} + (f_n^2 - f_{\delta}^2) \rho_p \rho_n^{10/3}] + \dfrac{6c_1^2}{M_n^4 M_p^2}[(f_n^2 - f_{\delta}^2) \rho_n^{8/3} \rho_p^{5/3} \nonumber \\
&+& (f_n - f_{\delta})^2 \rho_n^{5/3} \rho_p^{8/3}]
+ \dfrac{4c_1^2}{M_n^3 M_p^3}[(f_n^2 - f_{\delta}^2) \rho_n^{5/3} \rho_p^{8/3} + (f_n - f_{\delta})^2 \rho_n^{8/3} \rho_p^{5/3}] \nonumber \\
&+& 10c_2 \dfrac{\rho_n^{7/3}}{M_n^6}(-f_n \rho + f_{\delta} \rho_3)^2 -4c_1^2 f_{nl} \rho \left(\dfrac{\rho_n^{10/3}}{M_n^5} + \dfrac{\rho_n^{5/3} \rho_p^{5/3}}{M_p^2 M_n^3}\right) - 4 c_2 f_{nl}\dfrac{\rho^2 \rho_n^{7/3}}{M_n^5}\nonumber \\
&+& c_5 \dfrac{\rho_n^{13/3}}{M_n^{10}} + {\cal O}(\rho^{14/3})
\end{eqnarray}
In the same fashion the proton scalar density  reads:
\begin{eqnarray}
{\rho_s}_p&=& \rho_p +c_1 \dfrac{\rho_p^{5/3}}{M_p^2} + c_2 \dfrac{\rho_p^{7/3}}{M_p^4} +2c_1 \rho_p^{5/3}\left(\dfrac{f_n \rho +f_{\delta}\rho_3}{M_p^3}\right) + c_3 \dfrac{\rho_p^3}{M_p^6}\nonumber\\ 
&+& 2c_1^2 \rho_p^{5/3}f_n\left( \dfrac{\rho_n^{5/3}}{M_n^2 M_p^3}+\dfrac{\rho_p^{5/3}}{M_p^5}\right)+2c_1^2 \rho_n^{5/3}f_{\delta}\left( \dfrac{\rho_p^{5/3}}{M_p^5}-\dfrac{\rho_n^{5/3}}{M_n^2 M_p^3}\right)\nonumber\\ 
&+&4c_2 \rho_p^{7/3}\left(f_n\dfrac{\rho}{M_p^5} + f_{\delta}\dfrac{\rho_3}{M_p^5}\right)+c_4\dfrac{\rho_p^{11/3}}{M_p^8}\nonumber\\ 
&+&c_1\rho_p^{5/3}\left(\dfrac{-2f_{nl}\rho^2}{M_p^3}+\dfrac{3(-f_n \rho+f_\delta(\rho_n-\rho_p))^2}{M_p^4}\right)+ 2c_1 c_2 (f_n - f_{\delta})\frac{\rho_n^{7/3}\rho_p^{5/3}}{M_n^4 M_p^3}\nonumber \\
&+& 4c_1 c_2 (f_n - f_{\delta}) \frac{\rho_n^{5/3} \rho_p^ {7/3}}{M_n^ 2 M_p^ 5} +
6c_3 (f_n - f_{\delta}) \frac {\rho_n \rho_p^ 3}{M_p^7} + 6(c_1 c_ 2+c_3) (f_n + f_{\delta}) \frac{\rho_p^4}{M_p^ 7}\nonumber \\
&+& \dfrac{10c_1^2}{M_p^6}[(f_n + f_{\delta})^2 \rho_p^{13/3} + (f_n^2 - f_{\delta}^2) \rho_n \rho_p^{10/3}] + \dfrac{6c_1^2}{M_p^4 M_n^2}[(f_n^2 - f_{\delta}^2) \rho_p^{8/3} \rho_n^{5/3} \nonumber \\
&+& (f_n - f_{\delta})^2 \rho_p^{5/3} \rho_n^{8/3}]
+ \dfrac{4c_1^2}{M_p^3 M_n^3}[(f_n^2 - f_{\delta}^2) \rho_p^{5/3} \rho_n^{8/3} + (f_n - f_{\delta})^2 \rho_p^{8/3} \rho_n^{5/3}] \nonumber \\
&+& 10c_2 \dfrac{\rho_p^{7/3}}{M_p^6}(-f_n \rho - f_{\delta} \rho_3)^2 -4c_1^2 f_{nl} \rho \left(\dfrac{\rho_p^{10/3}}{M_p^5} + \dfrac{\rho_n^{5/3} \rho_p^{5/3}}{M_n^2 M_p^3}\right) - 4 c_2 f_{nl}\dfrac{\rho^2 \rho_p^{7/3}}{M_p^5} \nonumber \\
&+& c_5 \dfrac{\rho_p^{13/3}}{M_p^{10}} + {\cal O}(\rho^{14/3})
\end{eqnarray}

As in the $n-\Lambda$ matter, the expansion for the energy density is necessary and up to ${\cal O}(\rho^{13/3})$ the individual contributions read

\begin{eqnarray}
{\cal E}_{2}&=&\dfrac{1}{2}\left (\dfrac{1}{4}f_\rho - f_{\delta}\right )\rho_3^2+\dfrac{1}{2}(h_n-f_n)\rho^2
\end{eqnarray}
\begin{eqnarray}
{\cal E}_{7/3}&=&-\dfrac{c_2}{3}\left(\dfrac{\rho_n^{7/3}}{M_n^3} + \dfrac{\rho_p^{7/3}}{M_p^3}\right)
\end{eqnarray}
\begin{eqnarray}
{\cal E}_{8/3}&=&-c_1 \left[f_n \rho \left(\dfrac{\rho_n^{5/3}}{M_n^2}+\dfrac{\rho_p^{5/3}}{M_p^2}\right) -f_{\delta}\rho_3\left(\dfrac{\rho_n^{5/3}}{M_n^2}-\dfrac{\rho_p^{5/3}}{M_p^2}\right) \right]
\end{eqnarray}
\begin{eqnarray}
{\cal E}_{3}&=& \dfrac{1}{3}f_{nl}\rho^3 -\dfrac{1}{5}c_3 \left( \dfrac{\rho_n^3}{M_n^5} + \dfrac{\rho_p^3}{M_p^5} \right)
\end{eqnarray}
\begin{eqnarray}
{\cal E}_{10/3}&=&-\dfrac{1}{2}c_1^2 \left[f_{\delta}\left(\dfrac{(M_p^2 \rho_n^{5/3}-M_n^2 \rho_p^{5/3})^2}{M_n^4 M_p^4}\right)+f_{n}\left(\dfrac{(M_p^2 \rho_n^{5/3}+M_n^2 \rho_p^{5/3})^2}{M_n^4 M_p^4}\right)\right]\\ \nonumber &-&c_2\left[f_{\delta}\rho_3\left(\dfrac{\rho_p^{7/3}}{M_p^4}-\dfrac{\rho_n^{7/3}}{M_n^4}\right)+f_{n}\rho\left(\dfrac{\rho_p^{7/3}}{M_p^4}+\dfrac{\rho_n^{7/3}}{M_n^4}\right)\right]
\end{eqnarray}
\begin{eqnarray}
{\cal E}_{11/3}&=&-\dfrac{c_4}{7}\left(\dfrac{\rho_p^{11/3}}{M_p^7}+\dfrac{\rho_n^{11/3}}{M_n^7}\right)-c_1\left[-f_{nl}\rho^2\left(\dfrac{\rho_p^{5/3}}{M_p^2}+\dfrac{\rho_n^{5/3}}{M_n^2}\right)+f_{\delta}^2\rho_3^2\left(\dfrac{\rho_p^{5/3}}{M_p^3}+\dfrac{\rho_n^{5/3}}{M_n^3}\right)\right] \\ \nonumber
&-&c_1\left[f_n^2 \rho^2 \left(\dfrac{\rho_p^{5/3}}{M_p^3}+\dfrac{\rho_n^{5/3}}{M_n^3}\right) - 2f_n f_{\delta}(\rho_n^2 - \rho_p^2)\left(\dfrac{\rho_p^{5/3}}{M_p^3}-\dfrac{\rho_n^{5/3}}{M_n^3}\right)\right]
\end{eqnarray}

\begin{eqnarray}
{\cal E}_{4}&=&-(c_1 c_2+c_3)(f_{\delta} + f_n)\left(\dfrac{\rho_n^4}{M_n^6} + \dfrac{\rho_p^4}{M_p^6}\right) + \dfrac{1}{4}(f_q - f_m)(\rho_n^4 + \rho_p^4)+ \dfrac{3}{2}(f_q-f_m)(\rho_n^{2}\rho_{p}^2) \nonumber \\
&+&c_3 (f_{\delta} - f_n)\left(\dfrac{\rho_n^3\rho_p}{M_n^6} + \dfrac{\rho_p^3\rho_n}{M_p^6}\right)+ (f_q - f_m)(\rho_n^{3}\rho_{p} + \rho_n\rho_{p}^{3}) \nonumber \\
&+&c_1c_2 (f_{\delta} - f_n)\left(\dfrac{\rho_n^{7/3}\rho_{p}^{5/3}}{M_n^4 M_p^2} + \dfrac{\rho_n^{5/3}\rho_{p}^{7/3}}{M_n^2M_p^4}\right)
\end{eqnarray}

\begin{eqnarray}
{\cal E}_{13/3}&=& 2c_1^2 \left[(f_{\delta}^2 - f_n^2)\left(\dfrac{\rho_n^{8/3}\rho_p^{5/3}}{M_p^2 M_n^3} + \dfrac{\rho_n^{5/3}\rho_p^{8/3}}{M_p^3 M_n^2}\right) - (f_{\delta}-f_n)^2 \left(\dfrac{\rho_n^{8/3}\rho_p^{5/3}}{M_p^3 M_n^2} + \dfrac{\rho_n^{5/3}\rho_p^{8/3}}{M_p^2 M_n^3}\right)\right] \nonumber \\
&+&\dfrac{f_{nl}}{M_n^2 M_p^2}(\rho_n^{5/3}\rho_p^{8/3} + \rho_n^{8/3}\rho_p^{5/3}) - 2 c_2 (f_{\delta}-f_n)^2 \left(\dfrac{\rho_n^{7/3}\rho_p^{2}}{M_n^5} + \dfrac{\rho_n^{2}\rho_p^{7/3}}{M_p^5}\right) \nonumber \\
&+&c_2 f_{nl} \left(\dfrac{\rho_n^{7/3}\rho_p^{2}}{M_n^4} + \dfrac{\rho_n^{2}\rho_p^{7/3}}{M_p^4}\right)+ (4 c_2 + 2c_1^2) (f_{\delta}^2-f_n^2) \left(\dfrac{\rho_n\rho_p^{10/3}}{M_p^5} + \dfrac{\rho_n^{10/3}\rho_p}{M_n^5}\right) \nonumber \\
&+&(2c_2 + c_1^2) f_{nl} \left(\dfrac{\rho_n\rho_p^{10/3}}{M_p^4} + \dfrac{\rho_n^{10/3}\rho_p}{M_n^4}\right) - 2(c_2 + c_1^2) (f_{\delta}+f_n)^2 \left(\dfrac{\rho_p^{13/3}}{M_p^5} + \dfrac{\rho_n^{13/3}}{M_n^5}\right) \nonumber \\
&+&(c_2 + c_1^2) f_{nl} \left(\dfrac{\rho_p^{13/3}}{M_p^5} + \dfrac{\rho_n^{13/3}}{M_n^5}\right) - \dfrac{c_5}{9} \left(\dfrac{\rho_p^{13/3}}{M_p^9} + \dfrac{\rho_n^{13/3}}{M_n^9}\right) \nonumber \\
\end{eqnarray}

Once again, from the above expansions, one can appreciate that non linear contributions of order-$b$ and order-$a^2$, not considered in previous applications \cite {Margueron:2007jc}, are present in our ${\cal E}_{4}$ term.

\subsection{Numerical Results}

We now study the convergence of the expansion in $n-p$ matter. According to
tests performed in Ref. \cite{Dutra2014}, the best parametrization that
takes into account the $\delta$ meson is the NL$\delta$ set defined in
Refs. \cite{Liu2002,MP2004}. In this parametrization, the baryon masses are $M_p=M_n=939.3 \, {\rm MeV}$ while 
meson masses are given by $m_\sigma=550$ MeV, $m_\omega=783$ MeV,
$m_{\rho}=763$ MeV and $m_\delta=980$ MeV. The couplings are given by:
$g_\sigma=8.9586$, $g_\omega=9.2383$, $g_\rho=13.7258$ and $g_\delta=7.8528$, $a=0.033 g_\sigma^3$ and $b=-0.0048 g_\sigma^4$. Also, in analogy with the $\Lambda$ fraction defined previously, let us define
the proton fraction as $Y_p=\rho_p/\rho$.

\begin{figure}[h]
\begin{tabular}{cc}
\subfloat[]
{\includegraphics[width=0.5\textwidth]{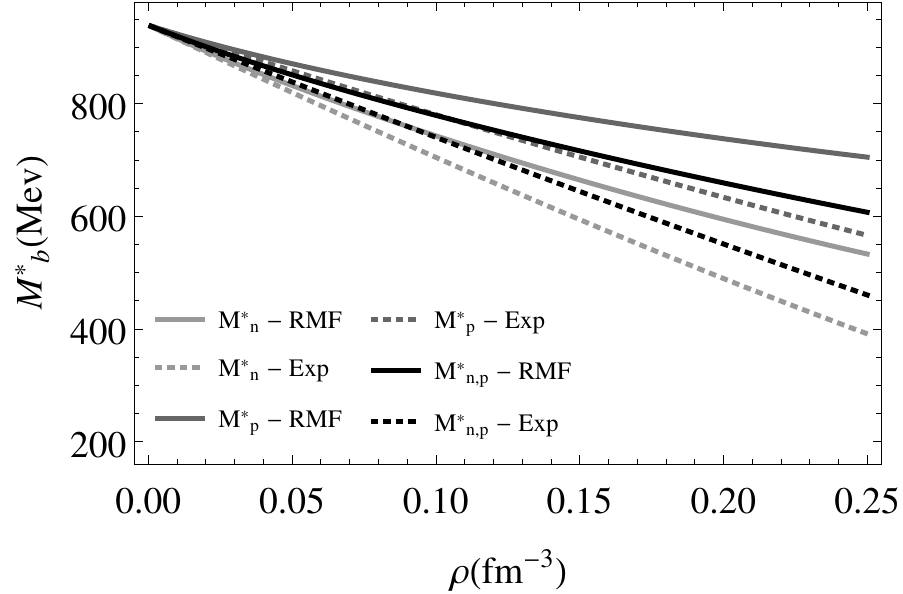}} &
\subfloat[]
{\includegraphics[width=0.5\textwidth]{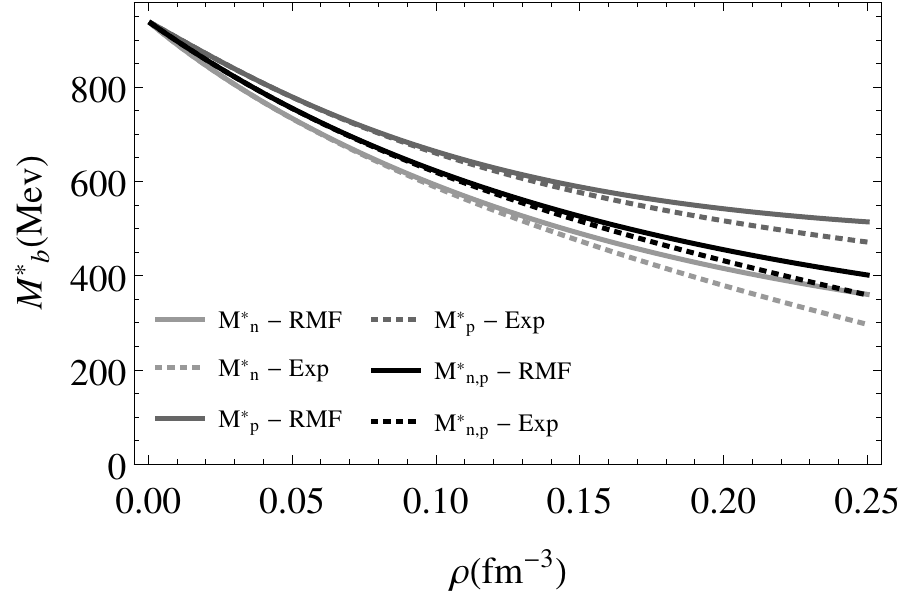}} \\
\end{tabular}
\caption{(Color online) Effective masses expanded up to order-$\rho^{10/3}$ compared with  (a) the RMF model and (b) the DDH$\delta$ model. For $Y_p=0.5$, the proton and neutron masses coincide ($M_{n,p}^*$) while  for
  $Y_p=0.1$ they are split due to the $\delta$ meson ($M_{n}^* \ne M_{p}^*$).}
\label{fig_mass_np}
\end{figure}

\begin{figure}[h]
	\begin{tabular}{cc}
		\subfloat[]
		{\includegraphics[width=0.5\textwidth]{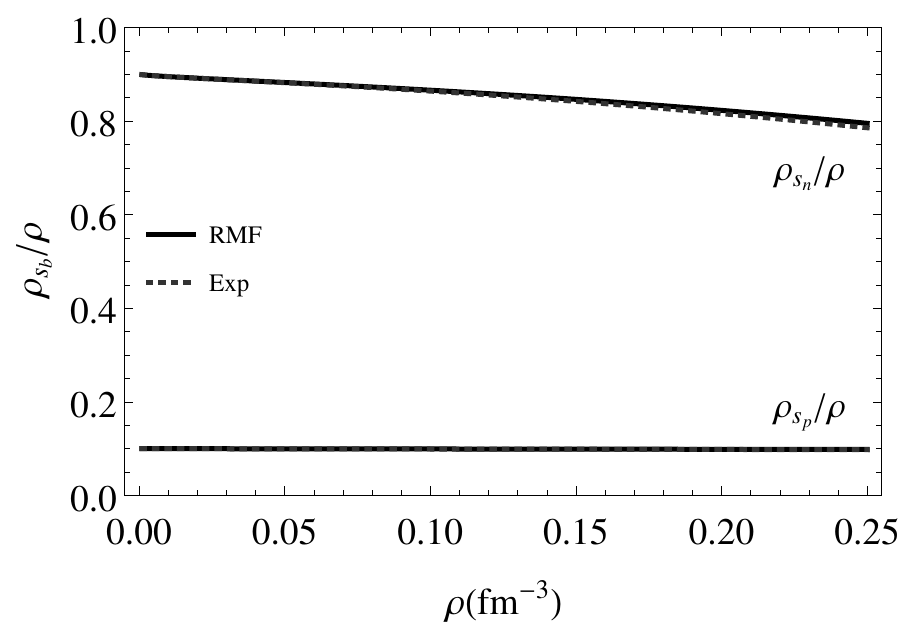}} &
		\subfloat[]
		{\includegraphics[width=0.5\textwidth]{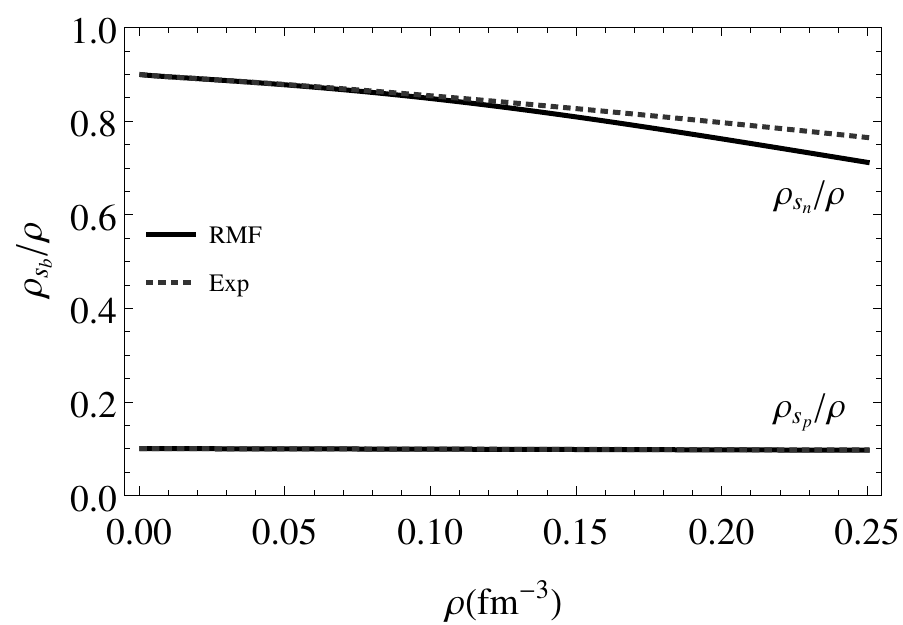}} \\
	\end{tabular}
	\caption{(Color online) Scalar density expanded up to order-$\rho^{13/3}$ compared with  the RMF model (a) and the DDH$\delta$ model (b) for the proton fraction $Y_p=0.5$.}
\label{fig_rhos_np}
\end{figure}

\begin{figure}[h]
\begin{tabular}{cc}
\subfloat[]
{\includegraphics[width=0.5\textwidth]{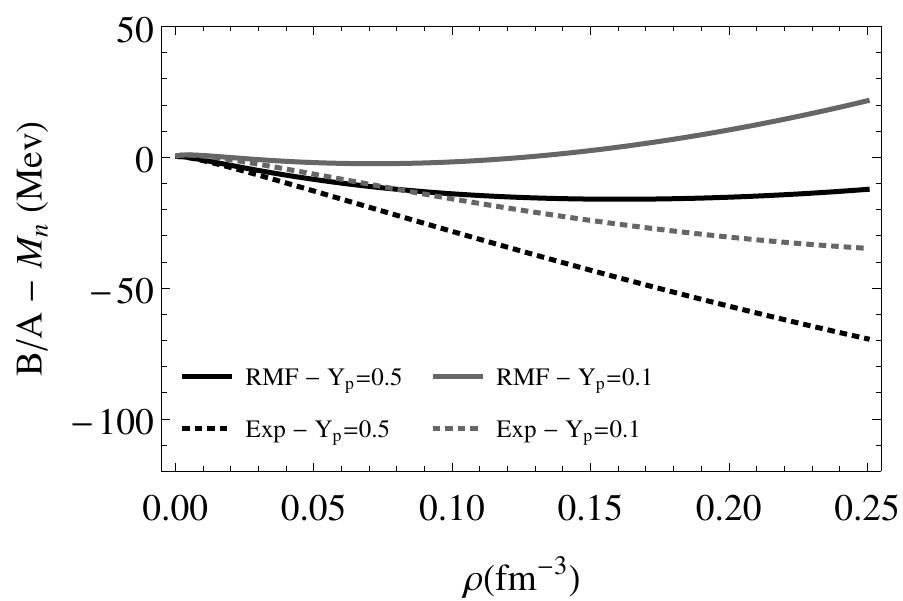}} &
\subfloat[]
{\includegraphics[width=0.5\textwidth]{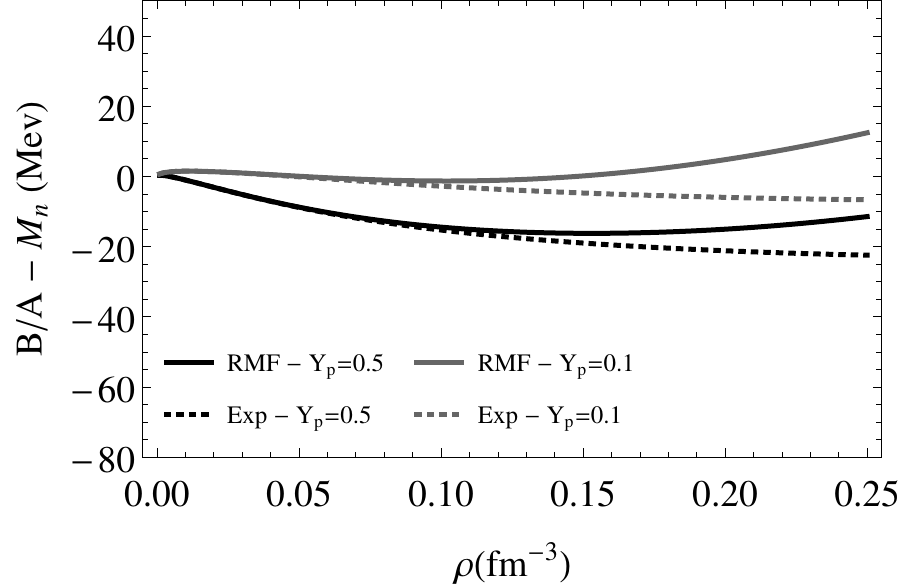}} \\
\end{tabular}
\caption{(Color online) Binding energy expanded up to order-$\rho^{13/3}$ compared with  the RMF model (a) and the DDH$\delta$ model (b) for proton fractions $Y_p=0.1$ (top curves) and $Y_p= 0.5$
  (bottom curves).}
\label{fig_binding_np}
\end{figure}

As in the $n-\Lambda$ case, we investigate the
convergence of the expansions by plotting the effective masses in Fig.
\ref{fig_mass_np}, the scalar densities in Fig. \ref{fig_rhos_np}   
and the binding energy in Fig. \ref{fig_binding_np} for two different
values of the proton fraction. Note that if we had plotted the energy density, the
results obtained in Fig. 2 of Ref. \cite{Margueron:2007jc} would be reproduced.

In this case, the density dependent model is the one that presents a better convergence for the effective masses, but not for the scalar densities. The fact that the chemical potentials strongly depend on the effective masses via the effective chemical potentials allows us to anticipate that the low density expansion  should give a better description of the spinodal boundary predicted by the density dependent model.  As far as the binding energies are concerned we cannot state that our results  show a considerable  improvement  in comparison with the results obtained in Ref. \cite{Providencia:2007dp} although the calculations presented here are more consistent since.

\begin{figure}[h]
\begin{tabular}{cc}
\subfloat[]
{\includegraphics[width=0.45\textwidth]{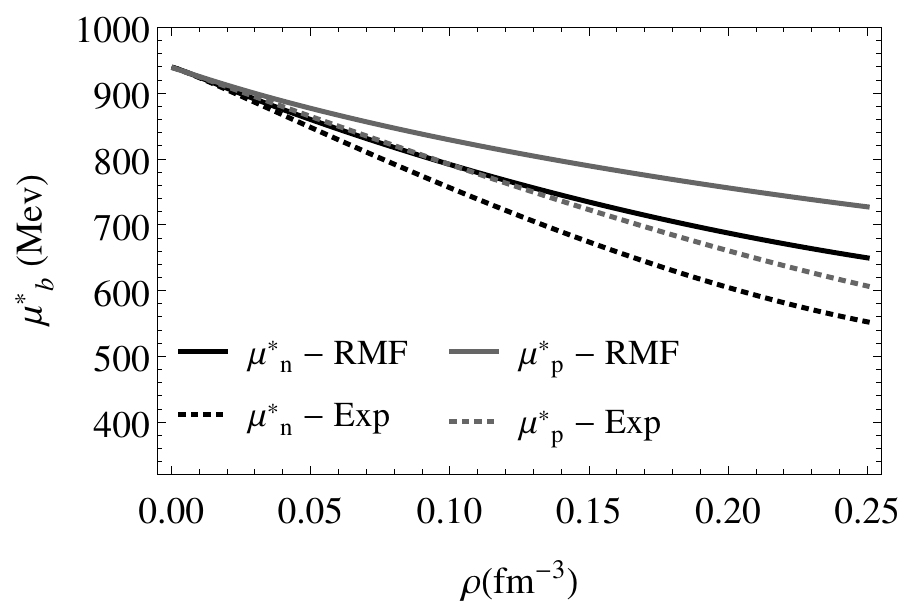}} &
\subfloat[]
{\includegraphics[width=0.45\textwidth]{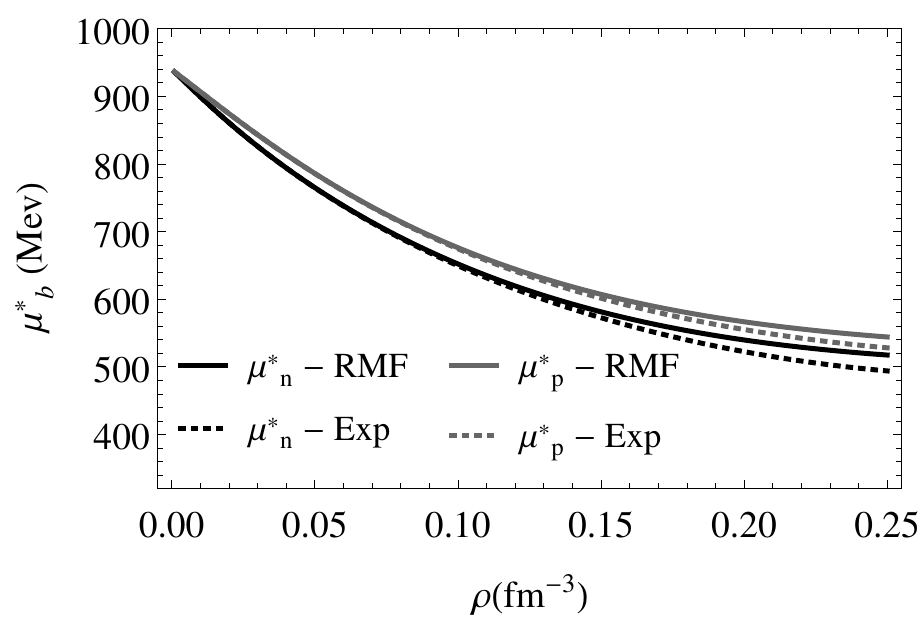}} \\
\subfloat[]
{\includegraphics[width=0.45\textwidth]{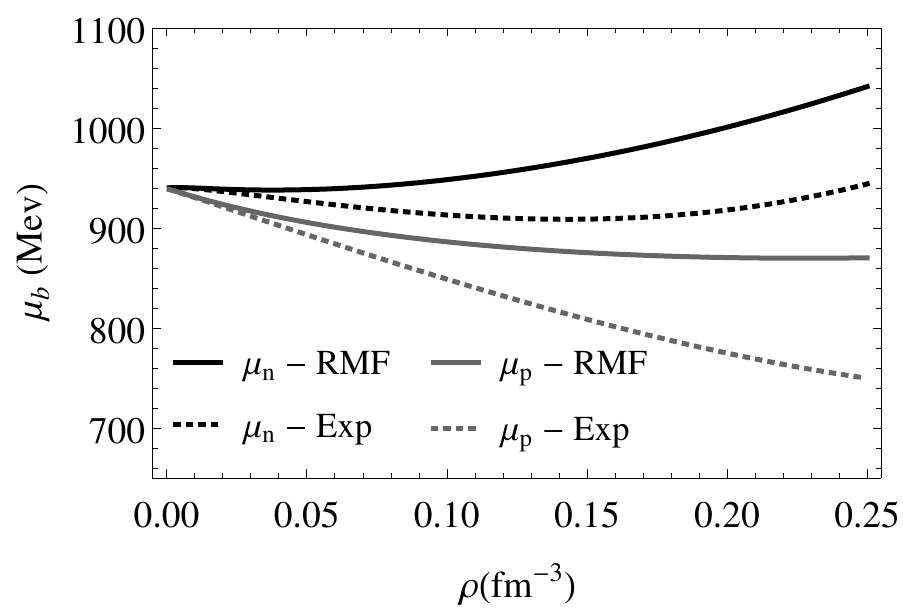}} &
\subfloat[]
{\includegraphics[width=0.45\textwidth]{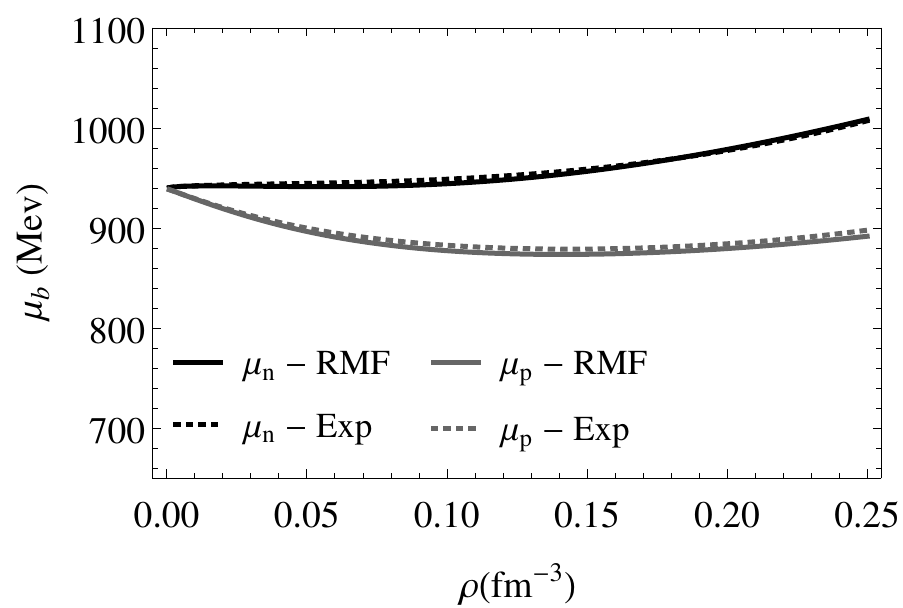}} \\
\end{tabular}
\caption{(Color online) Effective chemical for the RMF model (a) and the DDH$\delta$
  model (b). chemical potentials for for the RMF model (c) and the DDH$\delta$
  model (d). In both cases the proton fraction was set to $Y_p=0.1$ while the density expansion was performed up to order-$\rho^{13/3}$.}
\label{fig_nu_mu_np}
\end{figure}

We now turn our attention to the effective and real chemical
potentials, which are plotted in Fig.\ref{fig_nu_mu_np} for $Y_p=0.1$,
when the meson $\delta$ meson plays an important role. At this point it is
important to stress that the rearrangement term entering the chemical
potentials in the DDH$\delta$ model is also expanded in terms of the 
scalar densities when the low density expansion curves are displayed, which results 
in quite a good convergence, as compared with the results obtained from the RMF with fixed couplings. 

\begin{figure}[h]
\begin{tabular}{cc}
\subfloat[]
{\includegraphics[width=0.5\textwidth]{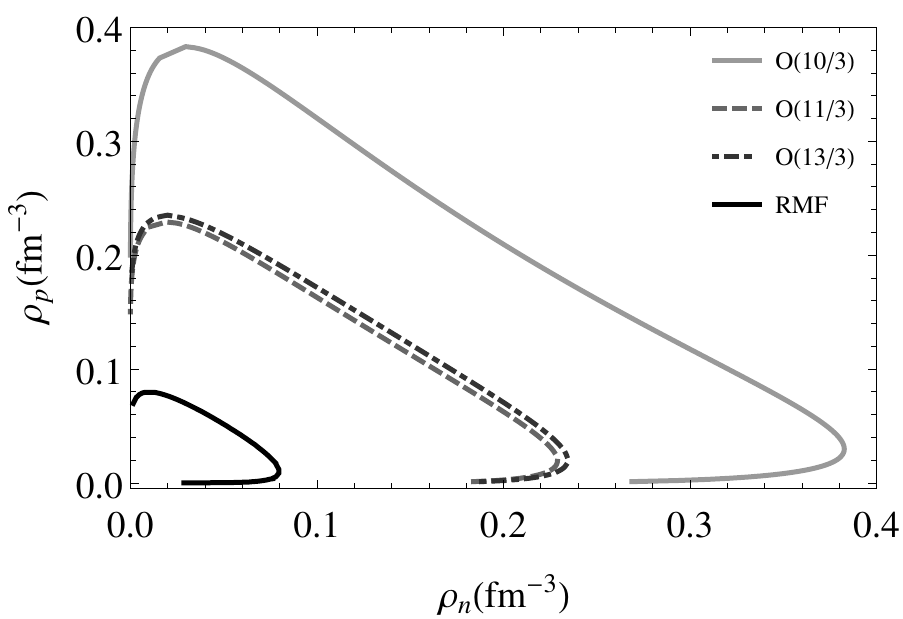}} &
\subfloat[]
{\includegraphics[width=0.5\textwidth]{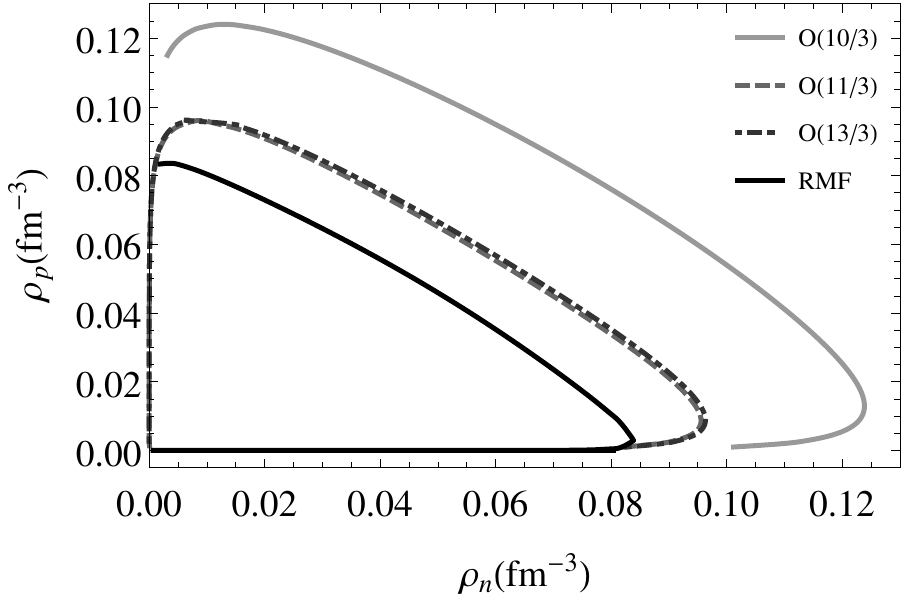}} \\
\end{tabular}
\caption{(Color online) Spinodal sections for (a) RMF and (b) DDH$\delta$  model. The expansion results shown are for orders $\rho^{10/3}$, $\rho^{11/3}$, and $\rho^{13/3}$.}
\label{fig_spinodalnp}
\end{figure}

The  spinodal sections, obtained from RMF
 and from the low density expansions are also plotted in Fig.\ref{fig_spinodalnp}.

Exactly as we have anticipated when examining the  effective masses  one observes that the slow convergence of the chemical potentials for the RMF model leads to a rather poor estimation of the spinodal zone when compared to the DDH$\delta$ model.  But even in the latter case  a sizeable difference appears in the spinodal borders even at order-$\rho^{13/3}$.  Nevertheless, our results represent a great improvement over the order-$\rho^{11/3}$ results obtained in Ref. \cite{Margueron:2007jc}.

\section{Conclusions}

In the present paper we have revisited the low density expansions of RMF models up to higher orders 
and in a more consistent way than the ones already existing in the literature. Both $n-\Lambda$ and $n-p$ matter were investigated with the help of a RMF model with fixed couplings and a density dependent coupling model. 
 We conclude that the isospin and strangeness degrees of freedom play quite different roles, based on the  observations described below. 

For $n-\Lambda$ matter, when we compare both models with their low density expansions, we have seen that the effective mass is very well described up to 0.25 fm$^{-3}$, but the $\Lambda$ scalar density is better described than the 
neutron scalar density.  When we check the binding energy expansions order by order, we see that they show a
convergent behavior, but when we look at the binding energy, we see that the the expansions stop converging before saturation density. Due to the new scale, we can see that what seems a small noise in the analyses of the energy density,
becomes a bigger problem. The chemical potentials, on the other hand, converge very well, but when we use their derivatives to obtain the spinodals, the DDH$\delta$ model is not reproduced due to the fact that the the rearrangement term depends on the scalar densities, not well reproduced by the expansions. Hence, our results indicate that the strange ($\Lambda$) sector is better described by the low expansion than the nucleonic ($n$) sector. Also, we observe that the spinodal contour predicted by our approximation is in excellent agreement with the one predicted by RMF when the couplings do not depend on the density.

For $n-p$ matter, the $\delta$ meson was incorporated and it seems to play an important role, so that the expansion is worse if asymmetric matter is considered. Despite of this problem we have been able to show that the poor results for the spinodal region obtained in Ref. \cite{Margueron:2007jc} can be significantly improved  by considering two higher orders terms within the low density expansion.
On the other hand, and contrary to our expectations, we could not make any considerable improvement in previous results for the binding energy published in Ref.  \cite{Providencia:2007dp}. 
As a global conclusion, we can say that an  expansion in powers of the density does not seem to provide a satisfactory mapping between relativistic and non-relativistic models  at any density for all physical quantities.
This result highlights the strong conceptual difference between an interaction picture and a meson exchange picture, with relativistic effects playing a role even at densities well below nuclear saturation. 

\section*{Acknowledgments}

This work was  partially supported by Capes(Brazil)/Cofecub (France),
joint international collaboration project number 853/15. DPM and MBP
acknowledge partial support also from CNPq (Brazil) and thank the
LPC-Caen for the hospitality. TO
acknowledges support from Capes (MSc scholarship). This work is also a 
part of the project CNPq-INCT-FNA Proc. No. 464898/2014-5.

\begin{appendix}

\section{Expansion of the rearrangement terms}

Up to order-$\rho^{13/3}$ the rearrangement terms for the $n-\Lambda$ case read

{\tiny \begin{eqnarray}
\Sigma^R_{n\Lambda}(\rho)&=&\frac{\partial\Gamma_{\omega n}} {\partial\rho}\left(\dfrac{\Gamma_{\omega n}}{m_\omega^2}\rho_n^2
+ \dfrac{\Gamma_{\omega\Lambda}}{m_\omega^2}\rho_n \rho_\Lambda\right)
+\frac{\partial\Gamma_{\omega\Lambda}} {\partial\rho}\left(\dfrac{\Gamma_{\omega n}}{m_\omega^2}\rho_n \rho_{\Lambda}
+ \dfrac{\Gamma_{\omega\Lambda}}{m_\omega^2}\rho_\Lambda^2\right)
+\frac{\partial\Gamma_\phi}{\partial\rho} \dfrac{\Gamma_\phi}{m_\phi^2}\rho_{\Lambda}^2 \nonumber \\
&-&\frac{\partial\Gamma_{\sigma n}}{\partial\rho}\left( \dfrac{\Gamma_{\sigma n}}{m_\sigma^2}\rho_n^2 + \dfrac{\Gamma_{\sigma\Lambda}}{m_\sigma^2}\rho_n \rho_{\Lambda}\right)
-\frac{\partial\Gamma_{\sigma\Lambda}} {\partial\rho}\left(\dfrac{\Gamma_{\sigma n}}{m_\sigma^2}\rho_n \rho_{\Lambda}
+ \dfrac{\Gamma_{\sigma\Lambda}}{m_\sigma^2}\rho_\Lambda^2\right) - \frac{\partial\Gamma_{\sigma^*\Lambda}}{\partial\rho} \dfrac{\Gamma_{\sigma^*\Lambda}}{m_{\sigma^*}^2}\rho_{\Lambda}^2 \nonumber \\
&-&\frac{c_1}{m_{\sigma}^2}\frac{\partial\Gamma_{\sigma n}} {\partial\rho}\left(\dfrac{2 \Gamma_{\sigma n}}{M_n^2}\rho_n^{8/3}
+ \dfrac{\Gamma_{\sigma\Lambda}}{M_n^2}\rho_n^{5/3} \rho_\Lambda + \dfrac{\Gamma_{\sigma\Lambda}}{M_{\Lambda}^2}\rho_n \rho_{\Lambda}^{5/3} \right) \nonumber \\
&-&\frac{c_1}{m_{\sigma}^2}\frac{\partial\Gamma_{\sigma \Lambda}} {\partial\rho}\left(\dfrac{2 \Gamma_{\sigma \Lambda}}{M_{\Lambda}^2}\rho_{\Lambda}^{8/3}
+ \dfrac{\Gamma_{\sigma n}}{M_n^2}\rho_n^{5/3} \rho_\Lambda + \dfrac{\Gamma_{\sigma n}}{M_{\Lambda}^2}\rho_n \rho_{\Lambda}^{5/3} \right) - \frac{2 c_1}{m_{\sigma^*}^2} \frac{\partial\Gamma_{\sigma^*\Lambda}}{\partial\rho} \dfrac{ \Gamma_{\sigma^*\Lambda}}{M_{\Lambda}^2}\rho_{\Lambda}^{8/3} \nonumber \\
&-&\frac{(c_1^2 + 2 c_2)}{m_{\sigma}^2} \left(\frac{\partial\Gamma_{\sigma n}} {\partial\rho}\dfrac{\Gamma_{\sigma n}}{M_{n}^4}\rho_n^{10/3}
+ \frac{\partial\Gamma_{\sigma \Lambda}} {\partial\rho}\dfrac{\Gamma_{\sigma \Lambda}}{M_{\Lambda}^4}\rho_{\Lambda}^{10/3}\right)- \frac{(c_1^2 + 2 c_2)}{m_{\sigma^*}^{2}} \left(\frac{\partial\Gamma_{\sigma^*\Lambda}} {\partial\rho}\dfrac{\Gamma_{\sigma^*\Lambda}}{M_{\Lambda}^4}\rho_{\Lambda}^{10/3}\right) \nonumber\\
&-&\frac{c_1^2}{m_{\sigma}^2}\left(\frac{\partial\Gamma_{\sigma n}} {\partial\rho}\dfrac{\Gamma_{\sigma \Lambda}}{M_{n}^2 M_{\Lambda}^2}
+ \frac{\partial\Gamma_{\sigma \Lambda}} {\partial\rho}\dfrac{\Gamma_{\sigma n}}{M_{n}^2 M_{\Lambda}^2}\right)\rho_n^{5/3} \rho_{\Lambda}^{5/3}\nonumber \\
&-&\frac{c_2}{m_{\sigma}^2}\left(\frac{\partial\Gamma_{\sigma n}} {\partial\rho}\Gamma_{\sigma \Lambda}
+ \frac{\partial\Gamma_{\sigma \Lambda}} {\partial\rho}\Gamma_{\sigma n}\right)\left(\dfrac{\rho_n^{7/3} \rho_{\Lambda}}{M_n^4} + \dfrac{\rho_n \rho_{\Lambda}^{7/3}}{M_{\Lambda}^4}\right)\nonumber \\   
&-&\frac{2c_1}{m_{\sigma}^2 M_n^3}(f_n \rho_n + f_{n\Lambda} \rho_{\Lambda}) \left[2\frac{\partial\Gamma_{\sigma n}} {\partial\rho}\Gamma_{\sigma n} \rho_n^{8/3}
+ \left(\frac{\partial\Gamma_{\sigma n}} {\partial\rho}\Gamma_{\sigma \Lambda} + \frac{\partial\Gamma_{\sigma \Lambda}} {\partial\rho}\Gamma_{\sigma n}\right)\rho_n^{5/3} \rho_{\Lambda} \right]\nonumber \\
&-&\frac{2c_1}{m_{\sigma}^2 M_{\Lambda}^3}(f_{n\Lambda} \rho_n + f_{\Lambda} \rho_{\Lambda}) \left[2\frac{\partial\Gamma_{\sigma \Lambda}} {\partial\rho}\Gamma_{\sigma \Lambda} \rho_{\Lambda}^{8/3}
+ \left(\frac{\partial\Gamma_{\sigma n}} {\partial\rho}\Gamma_{\sigma \Lambda} + \frac{\partial\Gamma_{\sigma \Lambda}} {\partial\rho}\Gamma_{\sigma n}\right)\rho_n \rho_{\Lambda}^{5/3} \right]\nonumber \\
&-&\frac{4c_1}{m_{\sigma^*}^2 M_{\Lambda}^3}(f_{n\Lambda} \rho_n + f_{\Lambda} \rho_{\Lambda})\frac{\partial\Gamma_{\sigma^*\Lambda}} {\partial\rho}\Gamma_{\sigma^*\Lambda} \rho_{\Lambda}^{8/3} - \frac{2 (c_1c_2 + c_3)}{m_{\sigma^*}^2} \frac{\partial\Gamma_{\sigma^* \Lambda}} {\partial\rho}\frac{\Gamma_{\sigma^* \Lambda}}{M_{\Lambda}^6}\rho_{\Lambda}^4 \nonumber \\
&-&\frac{2 (c_1c_2 + c_3)}{m_{\sigma}^2}\left( \frac{\partial\Gamma_{\sigma n}} {\partial\rho}\frac{\Gamma_{\sigma n}}{M_n^6}\rho_n^4 + \frac{\partial\Gamma_{\sigma \Lambda}} {\partial\rho}\frac{\Gamma_{\sigma \Lambda}}{M_{\Lambda}^6}\rho_{\Lambda}^4\right) \nonumber \\
&-&\left( \frac{\partial\Gamma_{\sigma n}} {\partial\rho}\frac{\Gamma_{\sigma \Lambda}}{m_{\sigma}^2} + \frac{\partial\Gamma_{\sigma \Lambda}}{\partial\rho}\frac{\Gamma_{\sigma n}}{m_{\sigma}^2}\right)\left[c_1c_2 \left(\frac{\rho_n^{7/3} \rho_{\Lambda}^{5/3}}{M_n^4 M_{\Lambda}^2} + \frac{\rho_n^{5/3} \rho_{\Lambda}^{7/3}}{M_n^2 M_{\Lambda}^4}\right) + c_3\left(\frac{\rho_n^3 \rho_{\Lambda}}{M_n^6} + \frac{\rho_n \rho_{\Lambda}^3}{M_{\Lambda}^6}\right) \right]\nonumber \\
&-&4\frac{\partial\Gamma_{\sigma n}} {\partial\rho}\frac{\Gamma_{\sigma n}}{m_{\sigma}^2}\left[2(c_1^2 + c_2) f_n\frac{\rho_n^{13/3}}{M_n^5} + (c_1^2 + 2 c_2) f_{n\Lambda}\frac{\rho_n^{10/3} \rho_{\Lambda}}{M_n^5}+ c_1^2 f_{n\Lambda}\frac{\rho_n^{8/3} \rho_{\Lambda}^{5/3}}{M_n^3 M_{\Lambda}^2}\right]\nonumber \\
&-&4\left(\frac{\partial\Gamma_{\sigma \Lambda}} {\partial\rho}\frac{\Gamma_{\sigma \Lambda}}{m_{\sigma}^2}+\frac{\partial\Gamma_{\sigma^* \Lambda}} {\partial\rho}\frac{\Gamma_{\sigma^* \Lambda}}{m_{\sigma^*}^2}\right)\left[2(c_1^2 + c_2) f_{\Lambda}\frac{\rho_{\Lambda}^{13/3}}{M_{\Lambda}^5} + (c_1^2 + 2 c_2) f_{n\Lambda}\frac{\rho_n \rho_{\Lambda}^{10/3}}{M_{\Lambda}^5}+ c_1^2 f_{n\Lambda}\frac{\rho_n^{5/3} \rho_{\Lambda}^{8/3}}{M_n^2 M_{\Lambda}^3}\right]\nonumber \\
&-& 2\left(\frac{\partial\Gamma_{\sigma \Lambda}} {\partial\rho}\frac{\Gamma_{\sigma n}}{m_{\sigma}^2}+\frac{\partial\Gamma_{\sigma n}} {\partial\rho}\frac{\Gamma_{\sigma \Lambda}}{m_{\sigma}^2}\right)\left[(c_1^2 + 2 c_2) \left(f_{n}\frac{\rho_n^{10/3} \rho_{\Lambda}}{M_{n}^5}+ f_{\Lambda}\frac{\rho_n \rho_{\Lambda}^{10/3}}{M_{\Lambda}^5}\right)+ c_1^2 \left(\frac{f_{\Lambda}}{M_n^2 M_{\Lambda}^3}+2\frac{f_{n\Lambda}}{M_n^3 M_{\Lambda}^2}\right)\rho_n^{5/3} \rho_{\Lambda}^{8/3}\right]\nonumber \\
&-& 2\left(\frac{\partial\Gamma_{\sigma \Lambda}} {\partial\rho}\frac{\Gamma_{\sigma n}}{m_{\sigma}^2}+\frac{\partial\Gamma_{\sigma n}} {\partial\rho}\frac{\Gamma_{\sigma \Lambda}}{m_{\sigma}^2}\right)\left[c_1^2 \left(\frac{f_{n}}{M_n^3 M_{\Lambda}^2}+2\frac{f_{n\Lambda}}{M_n^2 M_{\Lambda}^3}\right)\rho_n^{8/3} \rho_{\Lambda}^{5/3} + 2c_2 f_{n\Lambda} \left(\frac{\rho_n^{7/3} \rho_{\Lambda}^2}{M_{n}^5}+ \frac{\rho_n^2 \rho_{\Lambda}^{7/3}}{M_{\Lambda}^5}\right)\right]\nonumber \\
&+& O(\rho^{14/3})\,,
\end{eqnarray}}
since the order-$\rho^{3}$ is identically zero. The expanded rearrangement terms for the $n-p$ case can be obtained in a similar fashion yielding

{\tiny
\begin{eqnarray}
\Sigma^R_{np}(\rho)&=& \frac{\partial\Gamma_\omega} {\partial\rho}\dfrac{\Gamma_\omega}{m_\omega^2}\rho^2
+\frac{1}{4}\frac{\partial\Gamma_\rho}{\partial\rho} \dfrac{\Gamma_\rho}{m_\rho^2}\rho_3^2
-\frac{\partial\Gamma_\sigma}{\partial\rho} \dfrac{\Gamma_\sigma}{m_\sigma^2}\rho^2
-\frac{\partial\Gamma_\delta}{\partial\rho} \dfrac{\Gamma_\delta}{m_\delta^2}\rho_3^2 \nonumber \\
&-& 2 c_1\frac{\partial\Gamma_\delta}{\partial\rho} \dfrac{\Gamma_\delta}{m_\delta^2}\rho_3 \left(\dfrac{\rho_p^{5/3}}{M_p^2} - \dfrac{\rho_n^{5/3}}{M_n^2}\right) - 2 c_1 \frac{\partial\Gamma_\sigma}{\partial\rho} \dfrac{\Gamma_\sigma}{m_\sigma^2}\rho \left(\dfrac{\rho_p^{5/3}}{M_p^2} + \dfrac{\rho_n^{5/3}}{M_n^2}\right)\nonumber\\
&+& \frac{\partial\Gamma_\delta}{\partial\rho} \dfrac{\Gamma_\delta}{m_\delta^2} \left[-c_1^2\left(\dfrac{\rho_p^{5/3}}{M_p^2} - \dfrac{\rho_n^{5/3}}{M_n^2}\right)^2 - 2 c_2\rho_3\left(\dfrac{\rho_p^{7/3}}{M_p^4} - \dfrac{\rho_n^{7/3}}{M_n^4}\right)\right] \nonumber\\
&+& \frac{\partial\Gamma_\sigma}{\partial\rho} \dfrac{\Gamma_\sigma}{m_\sigma^2} \left[-c_1^2\left(\dfrac{\rho_p^{5/3}}{M_p^2} + \dfrac{\rho_n^{5/3}}{M_n^2}\right)^2 - 2 c_2\rho\left(\dfrac{\rho_p^{7/3}}{M_p^4} + \dfrac{\rho_n^{7/3}}{M_n^4}\right)\right] \nonumber\\
&-& 2\frac{\partial\Gamma_\delta}{\partial\rho} \dfrac{\Gamma_\delta}{m_\delta^2}\rho_3 \left[2c_1\dfrac{\rho_p^{5/3}}{M_p^3} (f_n \rho + f_{\delta}\rho_3) - 2 c_1\dfrac{\rho_n^{5/3}}{M_n^3}(f_n \rho - f_{\delta}\rho_3)\right] \nonumber\\
&-& 2\frac{\partial\Gamma_\sigma}{\partial\rho} \dfrac{\Gamma_\sigma}{m_\sigma^2}\rho \left[2c_1\dfrac{\rho_p^{5/3}}{M_p^3} (f_n \rho + f_{\delta}\rho_3) + 2 c_1\dfrac{\rho_n^{5/3}}{M_n^3}(f_n \rho - f_{\delta}\rho_3)\right] \nonumber\\
&+& \frac{\partial\Gamma_\delta}{\partial\rho} \dfrac{\Gamma_\delta}{m_\delta^2} \left[-2c_1c_2\left(\dfrac{\rho_p^{5/3}}{M_p^2} - \dfrac{\rho_n^{5/3}}{M_n^2}\right)\left(\dfrac{\rho_p^{7/3}}{M_p^4} - \dfrac{\rho_n^{7/3}}{M_n^4}\right) - 2 c_3\rho_3\left(\dfrac{\rho_p^{3}}{M_p^6} - \dfrac{\rho_n^3}{M_n^6}\right)\right] \nonumber\\
&+& \frac{\partial\Gamma_\sigma}{\partial\rho} \dfrac{\Gamma_\sigma}{m_\sigma^2} \left[-2c_1c_2\left(\dfrac{\rho_p^{5/3}}{M_p^2} + \dfrac{\rho_n^{5/3}}{M_n^2}\right)\left(\dfrac{\rho_p^{7/3}}{M_p^4} + \dfrac{\rho_n^{7/3}}{M_n^4}\right) - 2 c_3\rho\left(\dfrac{\rho_p^{3}}{M_p^6} + \dfrac{\rho_n^3}{M_n^6}\right)\right]\nonumber\\
&+& \frac{\partial\Gamma_\delta}{\partial\rho} \dfrac{\Gamma_\delta}{m_\delta^2} \left[-4c_1^2\left(\dfrac{\rho_p^{5/3}}{M_p^2} - \dfrac{\rho_n^{5/3}}{M_n^2}\right)\left(\dfrac{\rho_p^{5/3}}{M_p^3} (f_n \rho + f_{\delta}\rho_3) - \dfrac{\rho_n^{5/3}}{M_n^3}(f_n \rho - f_{\delta}\rho_3)\right)\right. \nonumber\\
&-& \left.4c_1^2\rho_3(f_{\delta} - f_n)\left(\dfrac{\rho_p^{5/3}\rho_n^{5/3}}{M_n^3 M_p^2}-\dfrac{\rho_p^{5/3}\rho_n^{5/3}}{M_p^3 M_n^2}\right) - 4c_1^2\rho_3(f_{\delta} + f_n)\left(\dfrac{\rho_p^{10/3}}{M_p^5} -\dfrac{\rho_n^{10/3}}{M_n^5}\right)\right.\nonumber\\
&-& \left.8c_2\rho_3\left(\dfrac{\rho_p^{7/3}}{M_p^5} (f_n \rho + f_{\delta}\rho_3) - \dfrac{\rho_n^{7/3}}{M_n^5}(f_n \rho - f_{\delta}\rho_3)\right)\right] \nonumber\\
&+& \frac{\partial\Gamma_\sigma}{\partial\rho} \dfrac{\Gamma_\sigma}{m_\sigma^2} \left[-4c_1^2\left(\dfrac{\rho_p^{5/3}}{M_p^2} + \dfrac{\rho_n^{5/3}}{M_n^2}\right)\left(\dfrac{\rho_p^{5/3}}{M_p^3} (f_n \rho + f_{\delta}\rho_3) + \dfrac{\rho_n^{5/3}}{M_n^3}(f_n \rho - f_{\delta}\rho_3)\right)\right. \nonumber\\
&+& \left.4c_1^2\rho(f_{\delta} - f_n)\left(\dfrac{\rho_p^{5/3}\rho_n^{5/3}}{M_n^3 M_p^2}+\dfrac{\rho_p^{5/3}\rho_n^{5/3}}{M_p^3 M_n^2}\right) - 4c_1^2\rho(f_{\delta} + f_n)\left(\dfrac{\rho_p^{10/3}}{M_p^5} +\dfrac{\rho_n^{10/3}}{M_n^5}\right)\right. \nonumber\\
&-& \left.8c_2\rho\left(\dfrac{\rho_p^{7/3}}{M_p^5} (f_n \rho + f_{\delta}\rho_3) + \dfrac{\rho_n^{7/3}}{M_n^5}(f_n \rho - f_{\delta}\rho_3)\right)\right] \nonumber\\
&+&{\cal O}(\rho^{14/3})
\end{eqnarray}}

\end{appendix}

\end{document}